\documentclass[aps,prl,twocolumn,groupedaddress]{revtex4-1}

\usepackage{graphicx}
\usepackage{amsmath}
\usepackage{color}
\usepackage[normalem]{ulem}
\usepackage[colorlinks=true,urlcolor=blue,citecolor=blue,linkcolor=blue]{hyperref}

\begin{document}


\title{Quantum metric and effective mass of a two-body bound state in a flat band}


\author{P. T{\"o}rm{\"a}}
\email[]{paivi.torma@aalto.fi}
\author{L. Liang}
\email[]{long.liang@aalto.fi}
\author{S. Peotta}
\affiliation{Department of Applied Physics, Aalto University, FI-00076 Aalto, Finland}


\newcommand{\PT}[1]{\textcolor{red}{{\bf PT: #1 }}}
\newcommand{\HAH}[1]{\textcolor{blue}{{\bf HAH: #1 }}}

\date{\today}

\begin{abstract}
We consider two-body bound states in a flat band of a multiband system. 
The existence of pair dispersion predicts the possibility of breaking the degeneracy of the band and creating order, such as superconductivity. Within a separable interaction potential approximation, we find that finiteness of the effective mass of a bound pair is determined by a band structure invariant, which in the uniform case becomes the quantum metric. The results offer a simple foundation to understand and predict flat band superconductivity. We propose an experiment to test the interaction-induced pair motion.
\end{abstract}

\maketitle

The concept of a flat band refers to Bloch bands of periodic systems, which are either perfectly dispersionless or where the band width is negligible compared to other energy scales. The effects of interactions and disorder are enhanced in such systems. This may lead to magnetic order~\cite{Lieb:1989,Tasaki:1992,Mielke:1993} and fractional Chern insulators~\cite{Bergholtz:2013}. Also high critical temperature for Cooper pairing has been predicted~\cite{Khodel:1990,Kopnin:2011,Heikkila:2011} in the case of attractive (effective) interactions. The group velocity of a single particle is zero and its effective mass $m_{eff}$ infinite in a flat band. The conventional single-band prediction for supercurrent, $n/m_{eff}$, where $n$ is the superfluid density, would thus suggest the absence of superfluidity. However, it has been predicted that, in a multiband system, interaction-induced movement of pairs is possible~\cite{Vidal:2000,Peotta:2015}. In recent experiments on bilayer graphene~\cite{Cao:2018,Yankowitz:2018}, superconductivity was found to coincide with the formation of flat bands at certain angles of the bilayer twist~\cite{Lopes:2007,Bistritzer:2011}. The large number of differing theoretical descriptions for these observations demonstrates the importance of understanding the origin of flat band superconductivity in as simple terms as possible. Here we show that \textit{the two-body problem} can be used to predict the possibility of superfluidity in a flat band. We find that the pair effective mass is characterized by band invariant quantities proportional to derivatives of the Bloch functions, in particular the quantum metric. 

In the Cooper problem~\cite{Cooper:1956}, the bound state energy of two fermions of opposite spins was solved while restricting the available phase space by the existence of the Fermi sea. This revealed that the Fermi sea is unstable towards formation of Cooper pairs for arbitrarily small attractive interaction, while without the Fermi sea, bound states require a finite interaction. 
Since flat band states are degenerate, a Fermi level cannot be defined for non-interacting particles. To form a many-body state with some symmetry broken order, such as superconductivity, the degeneracy has to be lifted. We now ask whether the tendency for breaking the degeneracy can be predicted from the two-body problem. In contrast to the Cooper-problem and conventional superconductivity, we are interested in the instability of the degeneracy instead of instability of the Fermi sea. In the flat band case, showing that a bound state exists in not as such sufficient: if the bound pair energy remains degenerate, then condensation to a certain pair momentum state --- the basic mechanism of superconductivity --- is not likely. We argue that the existence of a dispersion and finite effective mass for the pairs points to breaking the degeneracy and formation of superfluid/superconducting order in the many-body case. We now proceed to find  under which general conditions the flat band two-body problem in a multiband system may feature bound states with a dispersion. Our goal and results are different from the calculation of scattering states in a flat band~\cite{Valiente:2017}, and from the Cooper problem in a single dispersive band~\cite{Martikainen:2008}.      

We consider two interacting particles in a periodic potential and interacting via an interaction potential $\lambda V(1,2)$. The two-body Schr\"odinger equation is 
$\left[ T_1 + T_2 + \lambda V(1,2) \right] |\psi(1,2)\rangle = E |\psi(1,2)\rangle$, 
where $T_1 + T_2$ contains the kinetic energies and the periodic potential. The two particles can be either fermions or bosons.
The solution for $\lambda=0$ is the two-particle state given by
$(T_1 + T_2) |\varphi_n\rangle = E_n  |\varphi_n\rangle$,
where $n$ contains all quantum numbers (band index, lattice momentum, spin) of the two-particle state. Let $|\varphi_0(1,2)\rangle$ denote the state of the particles in absence of interactions and $E_0$ the corresponding energy. We consider a flat band where 
$E_n=E_0$ (or $E_n \simeq E_0$). We denote by $n$ the states in this flat band and by $n'$ those in other dispersive or flat bands. The solution that fulfills the Schr\"odinger equation 
is given by 
\begin{align}
|\psi (1,2)\rangle =& |\varphi_0(1,2)\rangle + \sum_{n\neq 0} \frac{|\varphi_n\rangle}{E-E_0} \langle \varphi_n | \lambda V| \psi(1,2)\rangle \nonumber \\
+& \sum_{n'} \frac{|\varphi_{n'}\rangle}{E-E_{n'}} \langle \varphi_{n'} | \lambda V| \psi(1,2)\rangle   \label{multibandSchrode} \\
E-E_0 =& \langle \varphi_0 | \lambda V| \psi(1,2)\rangle .
\end{align}
We then assume the isolated flat band limit: The lowest/highest energies $E'_{min/max}$ of the bands above/below the flat band are separated from it by a band gap that is larger than the interactions, $|E_0-E'_{min/max}|\ge|\lambda|$. If we further assume that $E$ is close to $E_0$ (weak interactions), then $|E-E'_{min/max}|\ge|\lambda|$ and the last term of Eq.~(\ref{multibandSchrode}) becomes negligible. 
We proceed with
\begin{align}
|\psi (1,2)\rangle =& |\varphi_0(1,2)\rangle + \frac{1}{E-E_0}\sum_{n\neq 0} |\varphi_n\rangle \langle \varphi_n | \lambda V| \psi(1,2)\rangle \nonumber \\
E_b &\equiv E-E_0 = \langle \varphi_0 | \lambda V| \psi(1,2)\rangle  , \label{Start2}
\end{align}
where $n$ refers to quantum numbers in the isolated flat band, and we introduced the notation $E_b$ for the pair binding energy. We use units where $\hbar$ and the system volume are set to one.

We use the Bloch functions $e^{i \mathbf{k} \cdot \mathbf{x}} m_\mathbf{k}(\mathbf{x})$ of the flat band where $\mathbf{k}$ is the lattice momentum and the band and spin indices are not marked explicitly. Then $\varphi_0(\mathbf{x}_1,\mathbf{x}_2) = e^{i \mathbf{k}_1 \cdot \mathbf{x}_1} m_{\mathbf{k}_1} (\mathbf{x}_1) 
e^{i \mathbf{k}_2 \cdot \mathbf{x}_2} m_{\mathbf{k}_2} (\mathbf{x}_2) 
= e^{i \mathbf{q} \cdot \mathbf{R}} e^{i \mathbf{k} \cdot \mathbf{r}} m_{\mathbf{k}+\frac{\mathbf{q}}{2}} (\mathbf{x}_1)  m_{-\mathbf{k}+\frac{\mathbf{q}}{2}} (\mathbf{x}_2)$, where $\mathbf{q}= \mathbf{k_1} + \mathbf{k_2}$, $\mathbf{k}= (\mathbf{k_1} - \mathbf{k_2})/2$, $\mathbf{R}= (\mathbf{x_1} + \mathbf{x_2})/2$, $\mathbf{r}= \mathbf{x_1} - \mathbf{x_2}$ are the center-of-mass (COM) and relative momenta and coordinates, respectively. 
We consider interaction potentials $V$, whose dependence on the COM coordinate has the same periodicity as the lattice, then the COM momentum $\mathbf{q}$ of the two particles is conserved. 
Even when we consider the two-body problem in the isolated flat band, the multiband nature of the system is inherent in the spatial dependence of the periodic part of the Bloch function, $m_\mathbf{k}(\mathbf{x})$. In the language of lattice models, it contains the orbital dependence of the Bloch function. 
We consider a general interaction potential $V=V(\mathbf{x}_1, \mathbf{x}_2)$ (instead of $V=V(\mathbf{r})$) to incorporate possible effects arising from the spatial (orbital) dependence of the interaction and the Bloch functions.

We now make our second approximation: consider a separable potential~\cite{Fetter_book} $V(\mathbf{x}_1, \mathbf{x}_2) \longrightarrow u(\mathbf{x}_1, \mathbf{x}_2)u(\mathbf{x}_1', \mathbf{x}_2')$, and assume $u$ real. 
The calculations are straightforward but, for convenience, intermediate steps are given in the Supplemental Material~\cite{Supplemental} (where we also show that an alternative approach using a variational ansatz produces the same results). The result becomes
\begin{equation}
E_b = \lambda \sum_{\mathbf{k}} |\tilde{u} (\mathbf{q},\mathbf{k})|^2   , \label{ApprResult}
\end{equation}
$\tilde{u} (\mathbf{q},\mathbf{k}) = \int d\textbf{x}_1 d\textbf{x}_2 e^{-i \mathbf{k} \cdot \mathbf{r}} m^*_{\mathbf{k}+\frac{\mathbf{q}}{2}} (\mathbf{x}_1)  m^*_{-\mathbf{k}+\frac{\mathbf{q}}{2}} (\mathbf{x}_2) 
u(\mathbf{x}_1, \mathbf{x}_2)$ 
and the momentum summation is over the first Brillouin zone.
This shows that, for attractive (effective) interactions ($\lambda<0$), a bound state ($E_b<0$) exists whenever $\tilde{u} (\mathbf{k_1},\mathbf{k_2})$ is non-zero for a sufficiently large number of momenta so that the sum in (\ref{ApprResult}) is non-vanishing in the thermodynamical limit. Importantly, the bound state energy is linearly proportional to the coupling constant $\lambda$. Bardeen-Cooper-Schrieffer (BCS) mean-field theory for a flat band system predicts linear dependence of the order parameter (pairing gap)~\cite{Miyahara:2007,Kopnin:2011} and the superfluid weight~\cite{Peotta:2015} on the coupling constant. Our result shows that this dependence is predicted already at the two-body level. The linear dependence is in striking contrast to the exponential suppression by $\lambda$ of the order parameter (pairing gap) in the BCS theory and two-body Cooper problem in a dispersive system. 

To find out whether the bound pair has a dispersion, we study $|\tilde{u} (\mathbf{k_1},\mathbf{k_2})|^2$ further. We bring back the original interaction potential using 
$u(\mathbf{x}_1, \mathbf{x}_2)u(\mathbf{x}_1', \mathbf{x}_2') \longrightarrow 
V(\mathbf{x}_1, \mathbf{x}_2) \delta(\mathbf{x}_1-\mathbf{x}_1') \delta(\mathbf{x}_2-\mathbf{x}_2')$. Furthermore, we assume contact interaction
$V(\mathbf{x}_1, \mathbf{x}_2)=V(\mathbf{x}_1) \delta(\mathbf{x}_1 - \mathbf{x}_2)$, consider interacting particles that have opposite spins, 
and use the relation (consequence of time reversal symmetry) $m^\uparrow_{\mathbf{k}}= m^{\downarrow *}_{-\mathbf{k}}\equiv m_{\mathbf{k}}$.
We obtain
\begin{equation}
E_b = \lambda \sum_{\mathbf{k}} \int d\mathbf{x} V(\mathbf{x}) | m_{\mathbf{k} + \frac{\mathbf{q}}{2}} (\mathbf{x}) m_{\mathbf{k} - \frac{\mathbf{q}}{2}} (\mathbf{x})|^2 
. \label{NonUnifFinal}
\end{equation}
The result is  intuitive: in a flat band, kinetic energy effects are absent, thus the pairing energy is given solely by the probability of the particles to overlap in space and the local interaction potential.

We now expand the result (\ref{NonUnifFinal}) with respect to small pair momentum, 
$m_{\mathbf{k} \pm \frac{\mathbf{q}}{2}} = m_{\mathbf{k}} \pm \frac{q_i}{2} \partial_i m_{\mathbf{k}}  + \frac{1}{8} q_i q_j \partial_i \partial_j m_{\mathbf{k}}  + \mathcal{O}(q^3)$. Summation is assumed over repeated indices and $\partial_i \equiv \partial/\partial k_i$.
This gives 
\begin{align}
& E_b  \simeq \lambda \sum_{\mathbf{k}} \int d\mathbf{x} V(\mathbf{x}) 
\left[ P_\mathbf{k}(\mathbf{x})^2  \nonumber \right. \\
&- \left. \frac{q_i q_j}{4}  \left( \partial_i P_\mathbf{k}(\mathbf{x}) \partial_j P_\mathbf{k}(\mathbf{x}) - P_\mathbf{k}(\mathbf{x}) \partial_i \partial_j P_\mathbf{k}(\mathbf{x})  
 \right)\right]  , \label{FinalEnergyBeforePartial} \\
&= \lambda \sum_{\mathbf{k}} \int d\mathbf{x} V(\mathbf{x}) 
\left[ P_\mathbf{k}(\mathbf{x})^2 
- \frac{q_i q_j}{2} \partial_i P_\mathbf{k}(\mathbf{x}) \partial_j P_\mathbf{k}(\mathbf{x}) 
\right]  ,   \label{FinalEnergy}
\end{align}
where $P_\mathbf{k}(\mathbf{x}) = m^*_{\mathbf{k}} (\mathbf{x}) m_{\mathbf{k}} (\mathbf{x}) = \langle \mathbf{x}|m_\mathbf{k}\rangle \langle m_\mathbf{k}|\mathbf{x}\rangle$ is the diagonal element of the Bloch state projector $P_\mathbf{k}=|m_\mathbf{k}\rangle \langle m_\mathbf{k}| $.
The effective mass tensor is therefore
\begin{align}
    \left[\frac{1}{m^*}\right]_{ij} = - \lambda \sum_{\mathbf{k}} &\int d\mathbf{x} V(\mathbf{x}) 
\left[\partial_i P_\mathbf{k}(\mathbf{x}) \partial_j P_\mathbf{k}(\mathbf{x})\right] .  \label{effMass}
\end{align}
This means that the existence of a finite, positive effective mass, and thus the possibility of breaking the degeneracy towards an ordered state, depends on the derivatives of the flat band projector in a simple way. 
The result is gauge invariant and independent of the basis. In the case of a trivial flat band, such as a single band lattice model with vanishing hopping, the periodic part of the Bloch function is independent of momentum and the pair mass remains infinite, preventing superfluidity. In a multiband lattice model, the derivatives can be finite. Bear in mind that the separable potential approximation leads to only one bound state, which is the sum of the exact bound states. 
However, we show in the following that, for several interesting lattice models, only one (significantly) \textit{dispersive} bound state exists and therefore the result (\ref{effMass}) is actually an excellent estimate for the pair effective mass, although the energy (\ref{FinalEnergy}) has an offset from the exact value. 

The superfluid weight $D^s_{ij}$, defined as the change in energy density $\delta E=D^s_{ij} q_i q_j/2 $ due to supercurrent $\mathbf{q}$, has been shown to be proportional to the \textit{quantum metric} by multiband mean-field theory~\cite{Peotta:2015,Iskin:2018a}, dynamical mean-field theory, density-matrix renormalization-group calculations and exact diagonalization~\cite{Julku:2016,Liang:2017a,Mondaini:2018} as well as by semiclassical~\cite{Liang:2017b} and perturbative~\cite{Tovmasyan:2016} approaches. The quantum metric~\cite{Provost:1980} (Fubini-Study metric) $g_{ij}(\mathbf{k})$ can be defined via the infinitesimal Bures distance between two quantum states $D_{Bures}^2=1-|\langle \psi(\mathbf{k}) | \psi(\mathbf{k}+d\mathbf{k})\rangle|^2 \simeq \sum_{ij} g_{ij}(\mathbf{k}) dk_i dk_j$ when a parameter $\mathbf{k}$ is varied. The quantum metric is the real part of the \textit{quantum geometric tensor} whose imaginary part is the Berry curvature. This connection allows us to determine the finite Chern number and Berry curvature as the lower bounds for superfluid weight using multiband BCS theory~\cite{Peotta:2015,Liang:2017a}. Remarkably, the essentials of such lower bounds can already be obtained from the two-body problem in a flat band, as we will show below.

To make a connection to previous many-body results, let us first note that the continuum results (\ref{NonUnifFinal})-(\ref{effMass}) can be mapped to a tight binding description (see Supplemental Material~\cite{Supplemental}), with the only consequence being that in (\ref{effMass}) the integration of position coordinate becomes a summation over the orbital coordinate within one unit cell ($u.c.$), and we write the system volume, namely the number of unit cells $N_c$, explicitly. We consider now an interaction potential that does not depend on position (orbital-independent potential), $V(\mathbf{x})=1$. The inverse effective mass (\ref{effMass}) then becomes $-\lambda/N_c \sum_{\mathbf{x} \in u.c.}  \langle \mathbf{x} | \partial_i P_\mathbf{k}|\mathbf{x}\rangle \langle \mathbf{x} | \partial_j P_\mathbf{k} |\mathbf{x} \rangle$. We approximate this by  $-\lambda/(N_c N_{orb}) \sum_{\mathbf{x},\mathbf{x}' \in u.c.} \langle \mathbf{x} | \partial_i P_\mathbf{k}|\mathbf{x'}\rangle \langle \mathbf{x'} | \partial_j P_\mathbf{k} |\mathbf{x} \rangle$, where $N_{orb}$ is the number of orbitals in the unit cell, which is valid when $\langle \mathbf{x} | \partial_i P_\mathbf{k}|\mathbf{x'}\rangle \sim \langle \mathbf{x} | \partial_i P_\mathbf{k}|\mathbf{x}\rangle$. The approximation is further motivated by its similarity to the BCS mean-field approach (see Supplemental Material~\cite{Supplemental}). Using $\sum_{\mathbf{x} \in u.c.} |\mathbf{x}\rangle \langle \mathbf{x}| = 1$ we obtain
\begin{align}
\left[\frac{1}{m^*}\right]_{ij} = \frac{-\lambda}{N_c N_{orb}} \sum_{\mathbf{k}} \mathrm {Tr} \left[ \partial_i P_\mathbf{k}\partial_j P_\mathbf{k}\right] = \frac{-\lambda}{N_c N_{orb}} \sum_{\mathbf{k}} g_{ij}(\mathbf{k}) ,  \label{qmetricresult}
\end{align}
where $g_{ij}(\mathbf{k})$ is the quantum metric. In Refs.~\cite{Peotta:2015,Liang:2017a,Tovmasyan:2016}, the superfluid weight $D^s_{ij}$ was derived in the isolated flat band approximation and assuming uniform pairing (precisely, orbital-independent interaction and $\int d\mathbf{k} |m_\mathbf{k}(\mathbf{x})|^2$ being the same for all $\mathbf{x}$). 
The relation between the superfluid weight and the effective mass in a flat band (see Supplemental Material~\cite{Supplemental}) is $D^s_{ij}\simeq n  (1/m^*)_{ij}$
when the Cooper pair density $n$ is small. 
Using this, the effective mass given by Ref.~\cite{Peotta:2015}
becomes $(1/m^*)_{ij} = - \lambda /(N_c N_{orb}) \sum_{\mathbf{k}} g_{ij}(\mathbf{k})$ which is the same as the result (\ref{qmetricresult}) obtained by the two-body calculation. 

The result (\ref{FinalEnergyBeforePartial}) inspires us to introduce and calculate the infinitesimal difference in local (orbital-specific) wavefunction overlaps as follows
\begin{align}
&D_{overlap} = |m_\mathbf{k}(\mathbf{x}) m_\mathbf{k}(\mathbf{x})|^2 - |m_{\mathbf{k}+d\mathbf{k}}(\mathbf{x})m_{\mathbf{k}-d\mathbf{k}}(\mathbf{x})|^2 \nonumber \\
&\simeq 
\sum_{ij}
\left[ \partial_i P_\mathbf{k}(\mathbf{x}) \partial_j P_\mathbf{k}(\mathbf{x}) - P_\mathbf{k}(\mathbf{x}) \partial_i \partial_j P_\mathbf{k}(\mathbf{x})  
 \right] dk_i
dk_j \nonumber \\ &\equiv \sum_{ij} g^{local}_{ij}(\mathbf{k},\mathbf{x}) dk_i
dk_j ,
\end{align}
where we have have defined the ``local quantum metric'' $g^{local}_{ij}(\mathbf{k},\mathbf{x})$, which is of the same form as the usual quantum metric but with the projector $P_\mathbf{k}=|m_\mathbf{k}\rangle \langle m_\mathbf{k}|$ replaced by its local matrix element  
$P_\mathbf{k}(\mathbf{x}) = \langle \mathbf{x}|m_\mathbf{k}\rangle \langle m_\mathbf{k}|\mathbf{x}\rangle = m^*_{\mathbf{k}} (\mathbf{x}) m_{\mathbf{k}} (\mathbf{x})$. The local quantum metric is both basis- and gauge-independent, unlike the conventional quantum metric and Berry curvature that are gauge-invariant but depend on the basis~\cite{Dobard:2015,Haruki:2018}, on the other hand, it is not positive semidefinite and thus not a Riemannian metric. We have shown in (\ref{FinalEnergyBeforePartial}) that the local quantum metric determines the flat band bound pair effective mass, and can be connected to the usual (global) quantum metric when assuming uniform pairing. Whether a physically meaningful ``local Berry connection (curvature)'' exists is a topic of future research. 

\begin{figure}
	 \includegraphics[width=\columnwidth]{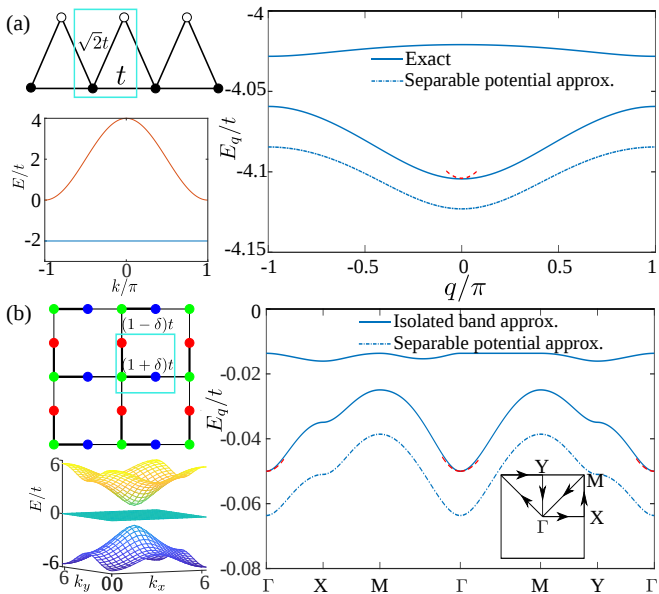}
	\caption{The energy dispersion of  two-body bound states for the sawtooth ladder (a) and Lieb lattice (b). Dashed lines show the result (\ref{FinalEnergy}) obtained by the separable potential approximation, and full lines the exact solution of the two-body problem for all bands (a) or for an isolated flat band (b). The lattice structures and non-interacting energy bands are shown on the left. For the Lieb lattice, the hopping integrals along the thick and thin links are $(1+\delta)t$ and $(1-\delta)t$ with $\delta=0.2$. The middle band is isolated from the other bands by a gap proportional to $\delta$. A nonzero $\delta$ breaks the four-fold rotational symmetry, and this is reflected in the two-body dispersion. The interaction strength is $\lambda=-0.2 t$ for the sawtooth ladder, and $\lambda=-0.1t$ for the Lieb lattice. The red curves show the quadratic dispersion with the effective mass given by \eqref{qmetricresult}. For the Lieb lattice, the uniform pairing condition is satisfied and the integrated quantum metric \eqref{qmetricresult} agrees very well with the numerical result. For the sawtooth ladder, the uniform pairing condition is violated, therefore the quantum metric approximation to the effective mass is not as good. }\label{Fig:two_body_dispersion}
\end{figure}

We test the analytical results against exact numerical solutions of the two-body Schr\"odinger equation in selected lattice models that feature flat bands. The contact interaction is used. 
For the 1D sawtooth ladder, which has one flat band (Fig.~\ref{Fig:two_body_dispersion}(a)), we solve the two-body problem numerically by taking into account all the bands. We find two bands formed by the bound states, and the dispersion of the bound state energy obtained from the separable potential approximation (\ref{FinalEnergy}) agrees (with an offset) with that of the exact lower band, see Fig.~\ref{Fig:two_body_dispersion}(a). 
 For the 2D Lieb lattice, the middle band is flat and can be made gapped from the lower and upper bands by staggered hopping. We solve the Lieb lattice two-body problem within the isolated band approximation by considering only the middle band. Again, we find that the lower bound state band dispersion agrees with the result from the separable potential approximation (\ref{FinalEnergy}). For results on the Harper model (Landau levels forming a flat band), see Supplemental Material~\cite{Supplemental}. 
 
\begin{figure}
\begin{tabular}{c}
\includegraphics[scale=0.7]{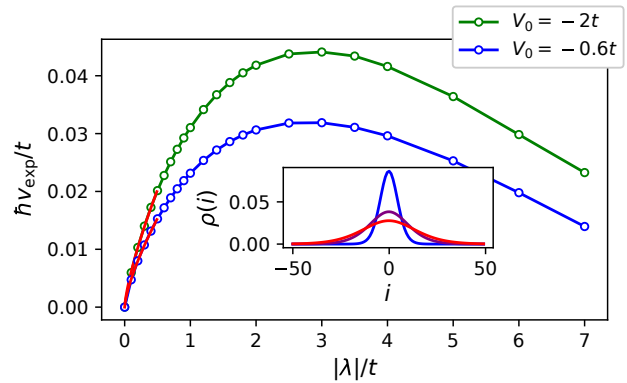}
\end{tabular}
\caption{Wave-packet expansion dynamics of a propagating two-body bound state in the sawtooth ladder. The initial wave-packet is obtained by calculating the ground state of a trapping potential of the form $V(i,\alpha) = V_0\cos(2\pi(i+b_\alpha)/N_{\rm c})$, with $b_{A}=0$, $b_{B} = 1/2$ and $N_{\rm c} = 200$ and then expressing it in terms of the propagating two-body bound states of the sawtooth ladder. The wave-packet is then released and expands as shown in the inset for the specific case $V_{0} = -2t$ and $\lambda = -3t$. There $i$ is the unit cell label and $\rho(i)$ the density distribution (identical for the two particles) summed over all orbitals in unit cell $i$, at times $100/t$, $250/t$ and $350/t$. In the main plot the expansion velocity $v_{\rm exp}$ is shown as a function of the interaction strength $|\lambda|$ for two different values of $V_0$. Red lines show fits to $|\lambda|^\gamma$, for details see the text.} \label{wavepacket}
\end{figure}

Our results can be directly tested in ultracold quantum gas experiments~\cite{Torma_book}. The propagation speeds of particles have been studied in experiments (see e.g.~\cite{Schneider:2012}) where the lattice potential is initially combined with a harmonic trapping potential that is later switched off, releasing the particles for motion. In a flat band, non-interacting atoms are expected to stay localized, while with interactions, pairs should propagate with a speed that increases linearly with the interaction strength and is essentially determined by Equation (\ref{effMass}). This is strikingly different from the dispersive band case where non-interacting particles propagate with a speed given by the hopping $t$ and pairs in the strong coupling limit with $t^2/\lambda$ velocity. Fig.~\ref{wavepacket} presents simulations of such an experiment for two particles.
The expansion velocity is obtained by fitting the free particle result $\langle \hat{x}^2(t)\rangle = \langle \hat{x}^2(0)\rangle+v_{\rm exp}^2t^2$ to the width $\langle \hat{x}^2(t)\rangle$ of the density distribution. The expansion velocity is controlled by the mass and the initial spread of the momentum distribution, since the effective mass approximation gives $v_{\rm exp} = \sqrt{\langle\, \hat{p}^2(0)\rangle}/m_{\rm eff}$. Using $\sqrt{\langle\, \hat{p}^2(0)\rangle}\propto |\lambda|^{-1/4}$ and $m_{\rm eff} \propto |\lambda|^{-1}$, one obtains $v_{\rm exp}\propto |\lambda|^{\gamma}$ with $\gamma = 0.75$. The value obtained by fitting of the data is $\gamma=0.74$ for $V_0=-2t$ and $\gamma= 0.72$ for $V_0=-0.3t$, in good agreement with the expected value. The $m_{\rm eff} \propto |\lambda|^{-1}$ behavior characteristic for a flat band is seen until $|\lambda|$ becomes comparable to the gap between the flat band and its neighboring bands (here $|\lambda| \simeq 2 t$). Experiments both for bosons and fermions would be interesting, although only the latter connects directly to superconductivity. By increasing the filling, an experiment could test to which extent the two-body predictions also describe the many-body case. Further, our results are symmetric in the coupling, and for $\lambda>0$, so-called repulsively bound pairs could be observed~\cite{Winkler2006_repulsivelyboundpairs}.

In summary, we show that the energy of a two-body bound state in a flat band is linearly proportional to the interaction constant and depends on the overlap of the periodic part of the Bloch functions and the orbital structure of the interaction potential in a simple way. The pair momentum dependence of the bound state energy can be used to determine the effective mass. Within the separable potential approximation, we find that it is essentially defined by a gauge- and basis-independent
quantity that we call the local quantum metric. With further approximations on the uniformity of the interactions and Bloch functions, we recover the dependence of the effective mass on geometric quantities, such as the (global) quantum metric and Berry curvature predicted earlier by many-body approaches. We demonstrate the adequacy of our approximate analytical results by comparison to exact solutions of the two-particle problem in the sawtooth ladder, Lieb lattice and Harper models; these and other flat band models can be realized for instance with ultracold gases~\cite{Aidelsburger:2013,Miyake:2013,Jotzu:2014,Flaeschner:2016,Ozawa:2017} or designer materials~\cite{Drost:2017,Slot:2017}, and the Brillouin zone integrated quantum metric can be measured~\cite{Asteria:2018}. We propose a direct signature of the predictions via an ultracold gas expansion experiment.  

Our results show that the two-particle problem already gives the salient features of the corresponding BCS mean-field (and other many-body) theory predictions. This suggests that in understanding and predicting superconductivity in flat bands of multiband systems, knowledge of the orbital dependence of the interaction and the non-interacting band structure can already be quite powerful. This may be advantageous when the single particle band structure alone is complex, for instance involving a large unit cell as in twisted bilayer 2D materials, and therefore formulating a suitable many-particle lattice model is a challenge. Our approach can easily be extended to different pairing symmetries. The two-body approach may also be, due to its computational lightness, well suited for materials discovery and optimization to find new flat band superconducting materials. The two-body effective mass also gives  the first order estimate of the Berezinskii-Kosterlitz-Thouless (BKT) temperature via the relation $T_{BKT} = \pi/2 \sqrt{\det[D^s(T_{BKT})]} \simeq \pi/2 \sqrt{\det[D^s(T=0)]} \simeq \pi n/2 \sqrt{\det[1/m^*]}$.
Furthermore, it gives a benchmark for full many-body descriptions to distinguish strong correlation effects. Our results highlight the role of the local and global quantum metric in a flat band and provide intuitive insight to the connection between superfluidity and quantum geometry. While all flat bands have a high density of states, the distances between the Bloch functions may differ in ways that are decisive for superconductivity.   

\emph{Acknowledgements.} This work was supported by the Academy of Finland under Projects No. 303351, No. 307419, No. 318987, and by the European Research Council (ERC-2013-AdG-340748-CODE). 

\bibliography{References}

\begin{thebibliography}{39}%
\makeatletter
\providecommand \@ifxundefined [1]{%
 \@ifx{#1\undefined}
}%
\providecommand \@ifnum [1]{%
 \ifnum #1\expandafter \@firstoftwo
 \else \expandafter \@secondoftwo
 \fi
}%
\providecommand \@ifx [1]{%
 \ifx #1\expandafter \@firstoftwo
 \else \expandafter \@secondoftwo
 \fi
}%
\providecommand \natexlab [1]{#1}%
\providecommand \enquote  [1]{``#1''}%
\providecommand \bibnamefont  [1]{#1}%
\providecommand \bibfnamefont [1]{#1}%
\providecommand \citenamefont [1]{#1}%
\providecommand \href@noop [0]{\@secondoftwo}%
\providecommand \href [0]{\begingroup \@sanitize@url \@href}%
\providecommand \@href[1]{\@@startlink{#1}\@@href}%
\providecommand \@@href[1]{\endgroup#1\@@endlink}%
\providecommand \@sanitize@url [0]{\catcode `\\12\catcode `\$12\catcode
  `\&12\catcode `\#12\catcode `\^12\catcode `\_12\catcode `\%12\relax}%
\providecommand \@@startlink[1]{}%
\providecommand \@@endlink[0]{}%
\providecommand \url  [0]{\begingroup\@sanitize@url \@url }%
\providecommand \@url [1]{\endgroup\@href {#1}{\urlprefix }}%
\providecommand \urlprefix  [0]{URL }%
\providecommand \Eprint [0]{\href }%
\providecommand \doibase [0]{http://dx.doi.org/}%
\providecommand \selectlanguage [0]{\@gobble}%
\providecommand \bibinfo  [0]{\@secondoftwo}%
\providecommand \bibfield  [0]{\@secondoftwo}%
\providecommand \translation [1]{[#1]}%
\providecommand \BibitemOpen [0]{}%
\providecommand \bibitemStop [0]{}%
\providecommand \bibitemNoStop [0]{.\EOS\space}%
\providecommand \EOS [0]{\spacefactor3000\relax}%
\providecommand \BibitemShut  [1]{\csname bibitem#1\endcsname}%
\let\auto@bib@innerbib\@empty
\bibitem [{\citenamefont {Lieb}(1989)}]{Lieb:1989}%
  \BibitemOpen
  \bibfield  {author} {\bibinfo {author} {\bibfnamefont {E.~H.}\ \bibnamefont
  {Lieb}},\ }\href {http://link.aps.org/doi/10.1103/PhysRevLett.62.1201}
  {\bibfield  {journal} {\bibinfo  {journal} {Phys. Rev. Lett.}\ }\textbf
  {\bibinfo {volume} {62}},\ \bibinfo {pages} {1201} (\bibinfo {year}
  {1989})}\BibitemShut {NoStop}%
\bibitem [{\citenamefont {Tasaki}(1992)}]{Tasaki:1992}%
  \BibitemOpen
  \bibfield  {author} {\bibinfo {author} {\bibfnamefont {H.}~\bibnamefont
  {Tasaki}},\ }\href {http://link.aps.org/doi/10.1103/PhysRevLett.69.1608}
  {\bibfield  {journal} {\bibinfo  {journal} {Phys. Rev. Lett.}\ }\textbf
  {\bibinfo {volume} {69}},\ \bibinfo {pages} {1608} (\bibinfo {year}
  {1992})}\BibitemShut {NoStop}%
\bibitem [{\citenamefont {Mielke}\ and\ \citenamefont
  {Tasaki}(1993)}]{Mielke:1993}%
  \BibitemOpen
  \bibfield  {author} {\bibinfo {author} {\bibfnamefont {A.}~\bibnamefont
  {Mielke}}\ and\ \bibinfo {author} {\bibfnamefont {H.}~\bibnamefont
  {Tasaki}},\ }\href {http://projecteuclid.org/euclid.cmp/1104254245}
  {\bibfield  {journal} {\bibinfo  {journal} {Commun. Math. Phys.}\ }\textbf
  {\bibinfo {volume} {158}},\ \bibinfo {pages} {341} (\bibinfo {year}
  {1993})}\BibitemShut {NoStop}%
\bibitem [{\citenamefont {Bergholtz}\ and\ \citenamefont
  {Liu}(2013)}]{Bergholtz:2013}%
  \BibitemOpen
  \bibfield  {author} {\bibinfo {author} {\bibfnamefont {E.~J.}\ \bibnamefont
  {Bergholtz}}\ and\ \bibinfo {author} {\bibfnamefont {Z.}~\bibnamefont
  {Liu}},\ }\href {\doibase 10.1142/S021797921330017X} {\bibfield  {journal}
  {\bibinfo  {journal} {Int. J. Mod. Phys. B}\ }\textbf {\bibinfo {volume}
  {27}},\ \bibinfo {pages} {1330017} (\bibinfo {year} {2013})}\BibitemShut
  {NoStop}%
\bibitem [{\citenamefont {Khodel}\ and\ \citenamefont
  {Shaginyan}(1990)}]{Khodel:1990}%
  \BibitemOpen
  \bibfield  {author} {\bibinfo {author} {\bibfnamefont {V.~A.}\ \bibnamefont
  {Khodel}}\ and\ \bibinfo {author} {\bibfnamefont {V.~R.}\ \bibnamefont
  {Shaginyan}},\ }\href {http://jetpletters.ac.ru/ps/1143/article_17312.shtml}
  {\bibfield  {journal} {\bibinfo  {journal} {JETP Lett.}\ }\textbf {\bibinfo
  {volume} {51}},\ \bibinfo {pages} {553} (\bibinfo {year} {1990})}\BibitemShut
  {NoStop}%
\bibitem [{\citenamefont {Kopnin}\ \emph {et~al.}(2011)\citenamefont {Kopnin},
  \citenamefont {Heikkil\"a},\ and\ \citenamefont {Volovik}}]{Kopnin:2011}%
  \BibitemOpen
  \bibfield  {author} {\bibinfo {author} {\bibfnamefont {N.~B.}\ \bibnamefont
  {Kopnin}}, \bibinfo {author} {\bibfnamefont {T.~T.}\ \bibnamefont
  {Heikkil\"a}}, \ and\ \bibinfo {author} {\bibfnamefont {G.~E.}\ \bibnamefont
  {Volovik}},\ }\href {http://link.aps.org/doi/10.1103/PhysRevB.83.220503}
  {\bibfield  {journal} {\bibinfo  {journal} {Phys. Rev. B}\ }\textbf {\bibinfo
  {volume} {83}},\ \bibinfo {pages} {220503} (\bibinfo {year}
  {2011})}\BibitemShut {NoStop}%
\bibitem [{\citenamefont {Heikkil{\"a}}\ \emph {et~al.}(2011)\citenamefont
  {Heikkil{\"a}}, \citenamefont {Kopnin},\ and\ \citenamefont
  {Volovik}}]{Heikkila:2011}%
  \BibitemOpen
  \bibfield  {author} {\bibinfo {author} {\bibfnamefont {T.~T.}\ \bibnamefont
  {Heikkil{\"a}}}, \bibinfo {author} {\bibfnamefont {N.~B.}\ \bibnamefont
  {Kopnin}}, \ and\ \bibinfo {author} {\bibfnamefont {G.~E.}\ \bibnamefont
  {Volovik}},\ }\href {http://dx.doi.org/10.1134/S0021364011150045} {\bibfield
  {journal} {\bibinfo  {journal} {JETP Lett.}\ }\textbf {\bibinfo {volume}
  {94}},\ \bibinfo {pages} {233} (\bibinfo {year} {2011})}\BibitemShut
  {NoStop}%
\bibitem [{\citenamefont {Vidal}\ \emph {et~al.}(2000)\citenamefont {Vidal},
  \citenamefont {Dou\ifmmode~\mbox{\c{c}}\else \c{c}\fi{}ot}, \citenamefont
  {Mosseri},\ and\ \citenamefont {Butaud}}]{Vidal:2000}%
  \BibitemOpen
  \bibfield  {author} {\bibinfo {author} {\bibfnamefont {J.}~\bibnamefont
  {Vidal}}, \bibinfo {author} {\bibfnamefont {B.}~\bibnamefont
  {Dou\ifmmode~\mbox{\c{c}}\else \c{c}\fi{}ot}}, \bibinfo {author}
  {\bibfnamefont {R.}~\bibnamefont {Mosseri}}, \ and\ \bibinfo {author}
  {\bibfnamefont {P.}~\bibnamefont {Butaud}},\ }\href {\doibase
  10.1103/PhysRevLett.85.3906} {\bibfield  {journal} {\bibinfo  {journal}
  {Phys. Rev. Lett.}\ }\textbf {\bibinfo {volume} {85}},\ \bibinfo {pages}
  {3906} (\bibinfo {year} {2000})}\BibitemShut {NoStop}%
\bibitem [{\citenamefont {Peotta}\ and\ \citenamefont
  {T\"orm\"a}(2015)}]{Peotta:2015}%
  \BibitemOpen
  \bibfield  {author} {\bibinfo {author} {\bibfnamefont {S.}~\bibnamefont
  {Peotta}}\ and\ \bibinfo {author} {\bibfnamefont {P.}~\bibnamefont
  {T\"orm\"a}},\ }\href {http://dx.doi.org/10.1038/ncomms9944} {\bibfield
  {journal} {\bibinfo  {journal} {Nat. Commun.}\ }\textbf {\bibinfo {volume}
  {6}},\ \bibinfo {pages} {8944} (\bibinfo {year} {2015})}\BibitemShut
  {NoStop}%
\bibitem [{\citenamefont {Cao}\ \emph {et~al.}(2018)\citenamefont {Cao},
  \citenamefont {Fatemi}, \citenamefont {Fang}, \citenamefont {Watanabe},
  \citenamefont {Taniguchi}, \citenamefont {Kaxiras},\ and\ \citenamefont
  {Jarillo-Herrero}}]{Cao:2018}%
  \BibitemOpen
  \bibfield  {author} {\bibinfo {author} {\bibfnamefont {Y.}~\bibnamefont
  {Cao}}, \bibinfo {author} {\bibfnamefont {V.}~\bibnamefont {Fatemi}},
  \bibinfo {author} {\bibfnamefont {S.}~\bibnamefont {Fang}}, \bibinfo {author}
  {\bibfnamefont {K.}~\bibnamefont {Watanabe}}, \bibinfo {author}
  {\bibfnamefont {T.}~\bibnamefont {Taniguchi}}, \bibinfo {author}
  {\bibfnamefont {E.}~\bibnamefont {Kaxiras}}, \ and\ \bibinfo {author}
  {\bibfnamefont {P.}~\bibnamefont {Jarillo-Herrero}},\ }\href
  {http://dx.doi.org/10.1038/nature26160} {\bibfield  {journal} {\bibinfo
  {journal} {Nature}\ }\textbf {\bibinfo {volume} {556}},\ \bibinfo {pages}
  {43} (\bibinfo {year} {2018})}\BibitemShut {NoStop}%
\bibitem [{\citenamefont {Yankowitz}\ \emph {et~al.}()\citenamefont
  {Yankowitz}, \citenamefont {Chen}, \citenamefont {Polshyn}, \citenamefont
  {Watanabe}, \citenamefont {Taniguchi}, \citenamefont {Graf}, \citenamefont
  {Young},\ and\ \citenamefont {Dean}}]{Yankowitz:2018}%
  \BibitemOpen
  \bibfield  {author} {\bibinfo {author} {\bibfnamefont {M.}~\bibnamefont
  {Yankowitz}}, \bibinfo {author} {\bibfnamefont {S.}~\bibnamefont {Chen}},
  \bibinfo {author} {\bibfnamefont {H.}~\bibnamefont {Polshyn}}, \bibinfo
  {author} {\bibfnamefont {K.}~\bibnamefont {Watanabe}}, \bibinfo {author}
  {\bibfnamefont {T.}~\bibnamefont {Taniguchi}}, \bibinfo {author}
  {\bibfnamefont {D.}~\bibnamefont {Graf}}, \bibinfo {author} {\bibfnamefont
  {A.}~\bibnamefont {Young}}, \ and\ \bibinfo {author} {\bibfnamefont
  {C.}~\bibnamefont {Dean}},\ }\href@noop {} {\ }\Eprint
  {http://arxiv.org/abs/1808:07865} {arXiv:1808:07865} \BibitemShut {NoStop}%
\bibitem [{\citenamefont {Lopes~dos Santos}\ \emph {et~al.}(2007)\citenamefont
  {Lopes~dos Santos}, \citenamefont {Peres},\ and\ \citenamefont
  {Castro~Neto}}]{Lopes:2007}%
  \BibitemOpen
  \bibfield  {author} {\bibinfo {author} {\bibfnamefont {J.~M.~B.}\
  \bibnamefont {Lopes~dos Santos}}, \bibinfo {author} {\bibfnamefont
  {N.~M.~R.}\ \bibnamefont {Peres}}, \ and\ \bibinfo {author} {\bibfnamefont
  {A.~H.}\ \bibnamefont {Castro~Neto}},\ }\href {\doibase
  10.1103/PhysRevLett.99.256802} {\bibfield  {journal} {\bibinfo  {journal}
  {Phys. Rev. Lett.}\ }\textbf {\bibinfo {volume} {99}},\ \bibinfo {pages}
  {256802} (\bibinfo {year} {2007})}\BibitemShut {NoStop}%
\bibitem [{\citenamefont {Bistritzer}\ and\ \citenamefont
  {MacDonald}(2011)}]{Bistritzer:2011}%
  \BibitemOpen
  \bibfield  {author} {\bibinfo {author} {\bibfnamefont {R.}~\bibnamefont
  {Bistritzer}}\ and\ \bibinfo {author} {\bibfnamefont {A.~H.}\ \bibnamefont
  {MacDonald}},\ }\href {\doibase 10.1073/pnas.1108174108} {\bibfield
  {journal} {\bibinfo  {journal} {Proceedings of the National Academy of
  Sciences}\ }\textbf {\bibinfo {volume} {108}},\ \bibinfo {pages} {12233}
  (\bibinfo {year} {2011})}\BibitemShut {NoStop}%
\bibitem [{\citenamefont {Cooper}(1956)}]{Cooper:1956}%
  \BibitemOpen
  \bibfield  {author} {\bibinfo {author} {\bibfnamefont {L.~N.}\ \bibnamefont
  {Cooper}},\ }\href {\doibase 10.1103/PhysRev.104.1189} {\bibfield  {journal}
  {\bibinfo  {journal} {Phys. Rev.}\ }\textbf {\bibinfo {volume} {104}},\
  \bibinfo {pages} {1189} (\bibinfo {year} {1956})}\BibitemShut {NoStop}%
\bibitem [{\citenamefont {Valiente}\ and\ \citenamefont
  {Zinner}(2017)}]{Valiente:2017}%
  \BibitemOpen
  \bibfield  {author} {\bibinfo {author} {\bibfnamefont {M.}~\bibnamefont
  {Valiente}}\ and\ \bibinfo {author} {\bibfnamefont {N.~T.}\ \bibnamefont
  {Zinner}},\ }\href {http://stacks.iop.org/0953-4075/50/i=6/a=064004}
  {\bibfield  {journal} {\bibinfo  {journal} {Journal of Physics B: Atomic,
  Molecular and Optical Physics}\ }\textbf {\bibinfo {volume} {50}},\ \bibinfo
  {pages} {064004} (\bibinfo {year} {2017})}\BibitemShut {NoStop}%
\bibitem [{\citenamefont {Martikainen}(2008)}]{Martikainen:2008}%
  \BibitemOpen
  \bibfield  {author} {\bibinfo {author} {\bibfnamefont {J.-P.}\ \bibnamefont
  {Martikainen}},\ }\href {\doibase 10.1103/PhysRevA.78.035602} {\bibfield
  {journal} {\bibinfo  {journal} {Phys. Rev. A}\ }\textbf {\bibinfo {volume}
  {78}},\ \bibinfo {pages} {035602} (\bibinfo {year} {2008})}\BibitemShut
  {NoStop}%
\bibitem [{\citenamefont {Fetter}\ and\ \citenamefont
  {Walecka}(2003)}]{Fetter_book}%
  \BibitemOpen
  \bibfield  {author} {\bibinfo {author} {\bibfnamefont {A.}~\bibnamefont
  {Fetter}}\ and\ \bibinfo {author} {\bibfnamefont {J.}~\bibnamefont
  {Walecka}},\ }\href@noop {} {\emph {\bibinfo {title} {Quantum Theory of
  Many-Particle Systems}}}\ (\bibinfo  {publisher} {Dover Publications, Inc.},\
  \bibinfo {year} {2003})\BibitemShut {NoStop}%
\bibitem [{Sup()}]{Supplemental}%
  \BibitemOpen
  \href@noop {} {\bibinfo  {journal} {See Supplemental Material at URL for
  intermediate steps of the calculations, for calculations by using a
  variational ansatz, and for numerical results for the Harper model}\
  }\BibitemShut {NoStop}%
\bibitem [{\citenamefont {Miyahara}\ \emph {et~al.}(2007)\citenamefont
  {Miyahara}, \citenamefont {Kusuta},\ and\ \citenamefont
  {Furukawa}}]{Miyahara:2007}%
  \BibitemOpen
\bibfield  {journal} {  }\bibfield  {author} {\bibinfo {author} {\bibfnamefont
  {S.}~\bibnamefont {Miyahara}}, \bibinfo {author} {\bibfnamefont
  {S.}~\bibnamefont {Kusuta}}, \ and\ \bibinfo {author} {\bibfnamefont
  {N.}~\bibnamefont {Furukawa}},\ }\href@noop {} {\bibfield  {journal}
  {\bibinfo  {journal} {Physica C: Superconductivity}\ }\textbf {\bibinfo
  {volume} {460}},\ \bibinfo {pages} {1145} (\bibinfo {year}
  {2007})}\BibitemShut {NoStop}%
\bibitem [{\citenamefont {Iskin}(2018)}]{Iskin:2018a}%
  \BibitemOpen
  \bibfield  {author} {\bibinfo {author} {\bibfnamefont {M.}~\bibnamefont
  {Iskin}},\ }\href {\doibase 10.1103/PhysRevA.97.033625} {\bibfield  {journal}
  {\bibinfo  {journal} {Phys. Rev. A}\ }\textbf {\bibinfo {volume} {97}},\
  \bibinfo {pages} {033625} (\bibinfo {year} {2018})}\BibitemShut {NoStop}%
\bibitem [{\citenamefont {Julku}\ \emph {et~al.}(2016)\citenamefont {Julku},
  \citenamefont {Peotta}, \citenamefont {Vanhala}, \citenamefont {Kim},\ and\
  \citenamefont {T\"orm\"a}}]{Julku:2016}%
  \BibitemOpen
  \bibfield  {author} {\bibinfo {author} {\bibfnamefont {A.}~\bibnamefont
  {Julku}}, \bibinfo {author} {\bibfnamefont {S.}~\bibnamefont {Peotta}},
  \bibinfo {author} {\bibfnamefont {T.~I.}\ \bibnamefont {Vanhala}}, \bibinfo
  {author} {\bibfnamefont {D.-H.}\ \bibnamefont {Kim}}, \ and\ \bibinfo
  {author} {\bibfnamefont {P.}~\bibnamefont {T\"orm\"a}},\ }\href
  {http://link.aps.org/doi/10.1103/PhysRevLett.117.045303} {\bibfield
  {journal} {\bibinfo  {journal} {Phys. Rev. Lett.}\ }\textbf {\bibinfo
  {volume} {117}},\ \bibinfo {pages} {045303} (\bibinfo {year}
  {2016})}\BibitemShut {NoStop}%
\bibitem [{\citenamefont {Liang}\ \emph
  {et~al.}(2017{\natexlab{a}})\citenamefont {Liang}, \citenamefont {Vanhala},
  \citenamefont {Peotta}, \citenamefont {Siro}, \citenamefont {Harju},\ and\
  \citenamefont {T\"orm\"a}}]{Liang:2017a}%
  \BibitemOpen
  \bibfield  {author} {\bibinfo {author} {\bibfnamefont {L.}~\bibnamefont
  {Liang}}, \bibinfo {author} {\bibfnamefont {T.~I.}\ \bibnamefont {Vanhala}},
  \bibinfo {author} {\bibfnamefont {S.}~\bibnamefont {Peotta}}, \bibinfo
  {author} {\bibfnamefont {T.}~\bibnamefont {Siro}}, \bibinfo {author}
  {\bibfnamefont {A.}~\bibnamefont {Harju}}, \ and\ \bibinfo {author}
  {\bibfnamefont {P.}~\bibnamefont {T\"orm\"a}},\ }\href {\doibase
  10.1103/PhysRevB.95.024515} {\bibfield  {journal} {\bibinfo  {journal} {Phys.
  Rev. B}\ }\textbf {\bibinfo {volume} {95}},\ \bibinfo {pages} {024515}
  (\bibinfo {year} {2017}{\natexlab{a}})}\BibitemShut {NoStop}%
\bibitem [{\citenamefont {Mondaini}\ \emph {et~al.}()\citenamefont {Mondaini},
  \citenamefont {Batrouni},\ and\ \citenamefont {B.}}]{Mondaini:2018}%
  \BibitemOpen
  \bibfield  {author} {\bibinfo {author} {\bibfnamefont {R.}~\bibnamefont
  {Mondaini}}, \bibinfo {author} {\bibfnamefont {G.~G.}\ \bibnamefont
  {Batrouni}}, \ and\ \bibinfo {author} {\bibfnamefont {G.}~\bibnamefont
  {B.}},\ }\href@noop {} {\enquote {\bibinfo {title} {Pairing and
  superconductivity in the flat band: Creutz lattice},}\ }\bibinfo {note}
  {\href{https://arxiv.org/abs/1805.09359}{arXiv:1805.09359}}\BibitemShut
  {NoStop}%
\bibitem [{\citenamefont {Liang}\ \emph
  {et~al.}(2017{\natexlab{b}})\citenamefont {Liang}, \citenamefont {Peotta},
  \citenamefont {Harju},\ and\ \citenamefont {T\"orm\"a}}]{Liang:2017b}%
  \BibitemOpen
  \bibfield  {author} {\bibinfo {author} {\bibfnamefont {L.}~\bibnamefont
  {Liang}}, \bibinfo {author} {\bibfnamefont {S.}~\bibnamefont {Peotta}},
  \bibinfo {author} {\bibfnamefont {A.}~\bibnamefont {Harju}}, \ and\ \bibinfo
  {author} {\bibfnamefont {P.}~\bibnamefont {T\"orm\"a}},\ }\href {\doibase
  10.1103/PhysRevB.96.064511} {\bibfield  {journal} {\bibinfo  {journal} {Phys.
  Rev. B}\ }\textbf {\bibinfo {volume} {96}},\ \bibinfo {pages} {064511}
  (\bibinfo {year} {2017}{\natexlab{b}})}\BibitemShut {NoStop}%
\bibitem [{\citenamefont {Tovmasyan}\ \emph {et~al.}(2016)\citenamefont
  {Tovmasyan}, \citenamefont {Peotta}, \citenamefont {T\"orm\"a},\ and\
  \citenamefont {Huber}}]{Tovmasyan:2016}%
  \BibitemOpen
  \bibfield  {author} {\bibinfo {author} {\bibfnamefont {M.}~\bibnamefont
  {Tovmasyan}}, \bibinfo {author} {\bibfnamefont {S.}~\bibnamefont {Peotta}},
  \bibinfo {author} {\bibfnamefont {P.}~\bibnamefont {T\"orm\"a}}, \ and\
  \bibinfo {author} {\bibfnamefont {S.~D.}\ \bibnamefont {Huber}},\ }\href
  {\doibase 10.1103/PhysRevB.94.245149} {\bibfield  {journal} {\bibinfo
  {journal} {Phys. Rev. B}\ }\textbf {\bibinfo {volume} {94}},\ \bibinfo
  {pages} {245149} (\bibinfo {year} {2016})}\BibitemShut {NoStop}%
\bibitem [{\citenamefont {Provost}\ and\ \citenamefont
  {Vallee}(1980)}]{Provost:1980}%
  \BibitemOpen
  \bibfield  {author} {\bibinfo {author} {\bibfnamefont {J.~P.}\ \bibnamefont
  {Provost}}\ and\ \bibinfo {author} {\bibfnamefont {G.}~\bibnamefont
  {Vallee}},\ }\href {\doibase 10.1007/BF02193559} {\bibfield  {journal}
  {\bibinfo  {journal} {Communications in Mathematical Physics}\ }\textbf
  {\bibinfo {volume} {76}},\ \bibinfo {pages} {289} (\bibinfo {year}
  {1980})}\BibitemShut {NoStop}%
\bibitem [{\citenamefont {Dobard\ifmmode \check{z}\else
  \v{z}\fi{}i\ifmmode~\acute{c}\else \'{c}\fi{}}\ \emph
  {et~al.}(2015)\citenamefont {Dobard\ifmmode \check{z}\else
  \v{z}\fi{}i\ifmmode~\acute{c}\else \'{c}\fi{}}, \citenamefont
  {Dimitrijevi\ifmmode~\acute{c}\else \'{c}\fi{}},\ and\ \citenamefont
  {Milovanovi\ifmmode~\acute{c}\else \'{c}\fi{}}}]{Dobard:2015}%
  \BibitemOpen
  \bibfield  {author} {\bibinfo {author} {\bibfnamefont {E.}~\bibnamefont
  {Dobard\ifmmode \check{z}\else \v{z}\fi{}i\ifmmode~\acute{c}\else
  \'{c}\fi{}}}, \bibinfo {author} {\bibfnamefont {M.}~\bibnamefont
  {Dimitrijevi\ifmmode~\acute{c}\else \'{c}\fi{}}}, \ and\ \bibinfo {author}
  {\bibfnamefont {M.~V.}\ \bibnamefont {Milovanovi\ifmmode~\acute{c}\else
  \'{c}\fi{}}},\ }\href {\doibase 10.1103/PhysRevB.91.125424} {\bibfield
  {journal} {\bibinfo  {journal} {Phys. Rev. B}\ }\textbf {\bibinfo {volume}
  {91}},\ \bibinfo {pages} {125424} (\bibinfo {year} {2015})}\BibitemShut
  {NoStop}%
\bibitem [{\citenamefont {Watanabe}\ and\ \citenamefont
  {Oshikawa}(2018)}]{Haruki:2018}%
  \BibitemOpen
  \bibfield  {author} {\bibinfo {author} {\bibfnamefont {H.}~\bibnamefont
  {Watanabe}}\ and\ \bibinfo {author} {\bibfnamefont {M.}~\bibnamefont
  {Oshikawa}},\ }\href {\doibase 10.1103/PhysRevX.8.021065} {\bibfield
  {journal} {\bibinfo  {journal} {Phys. Rev. X}\ }\textbf {\bibinfo {volume}
  {8}},\ \bibinfo {pages} {021065} (\bibinfo {year} {2018})}\BibitemShut
  {NoStop}%
\bibitem [{\citenamefont {T\"orm\"a}\ and\ \citenamefont
  {Sengstock}(2014)}]{Torma_book}%
  \BibitemOpen
  \bibfield  {author} {\bibinfo {author} {\bibfnamefont {P.}~\bibnamefont
  {T\"orm\"a}}\ and\ \bibinfo {author} {\bibfnamefont {K.}~\bibnamefont
  {Sengstock}},\ }\href@noop {} {\emph {\bibinfo {title} {Quantum Gas
  Experiments Exploring Many-Body States}}}\ (\bibinfo  {publisher} {Imperial
  College Press},\ \bibinfo {year} {2014})\BibitemShut {NoStop}%
\bibitem [{\citenamefont {Schneider}\ \emph {et~al.}(2012)\citenamefont
  {Schneider}, \citenamefont {Hackerm\"uller}, \citenamefont {Ronzheimer},
  \citenamefont {Will}, \citenamefont {Braun}, \citenamefont {Best},
  \citenamefont {Bloch}, \citenamefont {Demler}, \citenamefont {Mandt},
  \citenamefont {Rasch},\ and\ \citenamefont {Rosch}}]{Schneider:2012}%
  \BibitemOpen
  \bibfield  {author} {\bibinfo {author} {\bibfnamefont {U.}~\bibnamefont
  {Schneider}}, \bibinfo {author} {\bibfnamefont {L.}~\bibnamefont
  {Hackerm\"uller}}, \bibinfo {author} {\bibfnamefont {J.}~\bibnamefont
  {Ronzheimer}}, \bibinfo {author} {\bibfnamefont {S.}~\bibnamefont {Will}},
  \bibinfo {author} {\bibfnamefont {S.}~\bibnamefont {Braun}}, \bibinfo
  {author} {\bibfnamefont {T.}~\bibnamefont {Best}}, \bibinfo {author}
  {\bibfnamefont {I.}~\bibnamefont {Bloch}}, \bibinfo {author} {\bibfnamefont
  {E.}~\bibnamefont {Demler}}, \bibinfo {author} {\bibfnamefont
  {S.}~\bibnamefont {Mandt}}, \bibinfo {author} {\bibfnamefont
  {D.}~\bibnamefont {Rasch}}, \ and\ \bibinfo {author} {\bibfnamefont
  {A.}~\bibnamefont {Rosch}},\ }\href@noop {} {\bibfield  {journal} {\bibinfo
  {journal} {Nature Physics}\ }\textbf {\bibinfo {volume} {8}},\ \bibinfo
  {pages} {215} (\bibinfo {year} {2012})}\BibitemShut {NoStop}%
\bibitem [{\citenamefont {{Winkler}}\ \emph {et~al.}(2006)\citenamefont
  {{Winkler}}, \citenamefont {{Thalhammer}}, \citenamefont {{Lang}},
  \citenamefont {{Grimm}}, \citenamefont {{Hecker Denschlag}}, \citenamefont
  {{Daley}}, \citenamefont {{Kantian}}, \citenamefont {{B{\"u}chler}},\ and\
  \citenamefont {{Zoller}}}]{Winkler2006_repulsivelyboundpairs}%
  \BibitemOpen
  \bibfield  {author} {\bibinfo {author} {\bibfnamefont {K.}~\bibnamefont
  {{Winkler}}}, \bibinfo {author} {\bibfnamefont {G.}~\bibnamefont
  {{Thalhammer}}}, \bibinfo {author} {\bibfnamefont {F.}~\bibnamefont
  {{Lang}}}, \bibinfo {author} {\bibfnamefont {R.}~\bibnamefont {{Grimm}}},
  \bibinfo {author} {\bibfnamefont {J.}~\bibnamefont {{Hecker Denschlag}}},
  \bibinfo {author} {\bibfnamefont {A.~J.}\ \bibnamefont {{Daley}}}, \bibinfo
  {author} {\bibfnamefont {A.}~\bibnamefont {{Kantian}}}, \bibinfo {author}
  {\bibfnamefont {H.~P.}\ \bibnamefont {{B{\"u}chler}}}, \ and\ \bibinfo
  {author} {\bibfnamefont {P.}~\bibnamefont {{Zoller}}},\ }\href {\doibase
  10.1038/nature04918} {\bibfield  {journal} {\bibinfo  {journal} {\nat}\
  }\textbf {\bibinfo {volume} {441}},\ \bibinfo {pages} {853} (\bibinfo {year}
  {2006})}\BibitemShut {NoStop}%
\bibitem [{\citenamefont {Aidelsburger}\ \emph {et~al.}(2013)\citenamefont
  {Aidelsburger}, \citenamefont {Atala}, \citenamefont {Lohse}, \citenamefont
  {Barreiro}, \citenamefont {Paredes},\ and\ \citenamefont
  {Bloch}}]{Aidelsburger:2013}%
  \BibitemOpen
  \bibfield  {author} {\bibinfo {author} {\bibfnamefont {M.}~\bibnamefont
  {Aidelsburger}}, \bibinfo {author} {\bibfnamefont {M.}~\bibnamefont {Atala}},
  \bibinfo {author} {\bibfnamefont {M.}~\bibnamefont {Lohse}}, \bibinfo
  {author} {\bibfnamefont {J.~T.}\ \bibnamefont {Barreiro}}, \bibinfo {author}
  {\bibfnamefont {B.}~\bibnamefont {Paredes}}, \ and\ \bibinfo {author}
  {\bibfnamefont {I.}~\bibnamefont {Bloch}},\ }\href
  {http://link.aps.org/doi/10.1103/PhysRevLett.111.185301} {\bibfield
  {journal} {\bibinfo  {journal} {Phys. Rev. Lett.}\ }\textbf {\bibinfo
  {volume} {111}},\ \bibinfo {pages} {185301} (\bibinfo {year}
  {2013})}\BibitemShut {NoStop}%
\bibitem [{\citenamefont {Miyake}\ \emph {et~al.}(2013)\citenamefont {Miyake},
  \citenamefont {Siviloglou}, \citenamefont {Kennedy}, \citenamefont {Burton},\
  and\ \citenamefont {Ketterle}}]{Miyake:2013}%
  \BibitemOpen
  \bibfield  {author} {\bibinfo {author} {\bibfnamefont {H.}~\bibnamefont
  {Miyake}}, \bibinfo {author} {\bibfnamefont {G.~A.}\ \bibnamefont
  {Siviloglou}}, \bibinfo {author} {\bibfnamefont {C.~J.}\ \bibnamefont
  {Kennedy}}, \bibinfo {author} {\bibfnamefont {W.~C.}\ \bibnamefont {Burton}},
  \ and\ \bibinfo {author} {\bibfnamefont {W.}~\bibnamefont {Ketterle}},\
  }\href {http://link.aps.org/doi/10.1103/PhysRevLett.111.185302} {\bibfield
  {journal} {\bibinfo  {journal} {Phys. Rev. Lett.}\ }\textbf {\bibinfo
  {volume} {111}},\ \bibinfo {pages} {185302} (\bibinfo {year}
  {2013})}\BibitemShut {NoStop}%
\bibitem [{\citenamefont {Jotzu}\ \emph {et~al.}(2014)\citenamefont {Jotzu},
  \citenamefont {Messer}, \citenamefont {Desbuquois}, \citenamefont {Lebrat},
  \citenamefont {Uehlinger}, \citenamefont {Greif},\ and\ \citenamefont
  {Esslinger}}]{Jotzu:2014}%
  \BibitemOpen
  \bibfield  {author} {\bibinfo {author} {\bibfnamefont {G.}~\bibnamefont
  {Jotzu}}, \bibinfo {author} {\bibfnamefont {M.}~\bibnamefont {Messer}},
  \bibinfo {author} {\bibfnamefont {R.}~\bibnamefont {Desbuquois}}, \bibinfo
  {author} {\bibfnamefont {M.}~\bibnamefont {Lebrat}}, \bibinfo {author}
  {\bibfnamefont {T.}~\bibnamefont {Uehlinger}}, \bibinfo {author}
  {\bibfnamefont {D.}~\bibnamefont {Greif}}, \ and\ \bibinfo {author}
  {\bibfnamefont {T.}~\bibnamefont {Esslinger}},\ }\href
  {http://dx.doi.org/10.1038/nature13915} {\bibfield  {journal} {\bibinfo
  {journal} {Nature}\ }\textbf {\bibinfo {volume} {515}},\ \bibinfo {pages}
  {237} (\bibinfo {year} {2014})}\BibitemShut {NoStop}%
\bibitem [{\citenamefont {Fl{\"a}schner}\ \emph {et~al.}(2016)\citenamefont
  {Fl{\"a}schner}, \citenamefont {Rem}, \citenamefont {Tarnowski},
  \citenamefont {Vogel}, \citenamefont {L\"uhmann}, \citenamefont {Sengstock},\
  and\ \citenamefont {Weitenberg}}]{Flaeschner:2016}%
  \BibitemOpen
  \bibfield  {author} {\bibinfo {author} {\bibfnamefont {N.}~\bibnamefont
  {Fl{\"a}schner}}, \bibinfo {author} {\bibfnamefont {B.~S.}\ \bibnamefont
  {Rem}}, \bibinfo {author} {\bibfnamefont {M.}~\bibnamefont {Tarnowski}},
  \bibinfo {author} {\bibfnamefont {D.}~\bibnamefont {Vogel}}, \bibinfo
  {author} {\bibfnamefont {D.-S.}\ \bibnamefont {L\"uhmann}}, \bibinfo {author}
  {\bibfnamefont {K.}~\bibnamefont {Sengstock}}, \ and\ \bibinfo {author}
  {\bibfnamefont {C.}~\bibnamefont {Weitenberg}},\ }\href
  {http://science.sciencemag.org/content/352/6289/1091.abstract} {\bibfield
  {journal} {\bibinfo  {journal} {Science}\ }\textbf {\bibinfo {volume}
  {352}},\ \bibinfo {pages} {1091} (\bibinfo {year} {2016})}\BibitemShut
  {NoStop}%
\bibitem [{\citenamefont {Ozawa}\ \emph {et~al.}(2017)\citenamefont {Ozawa},
  \citenamefont {Taie}, \citenamefont {Ichinose},\ and\ \citenamefont
  {Takahashi}}]{Ozawa:2017}%
  \BibitemOpen
  \bibfield  {author} {\bibinfo {author} {\bibfnamefont {H.}~\bibnamefont
  {Ozawa}}, \bibinfo {author} {\bibfnamefont {S.}~\bibnamefont {Taie}},
  \bibinfo {author} {\bibfnamefont {T.}~\bibnamefont {Ichinose}}, \ and\
  \bibinfo {author} {\bibfnamefont {Y.}~\bibnamefont {Takahashi}},\ }\href
  {\doibase 10.1103/PhysRevLett.118.175301} {\bibfield  {journal} {\bibinfo
  {journal} {Phys. Rev. Lett.}\ }\textbf {\bibinfo {volume} {118}},\ \bibinfo
  {pages} {175301} (\bibinfo {year} {2017})}\BibitemShut {NoStop}%
\bibitem [{\citenamefont {Drost}\ \emph {et~al.}(2017)\citenamefont {Drost},
  \citenamefont {Ojanen}, \citenamefont {Harju},\ and\ \citenamefont
  {Liljeroth}}]{Drost:2017}%
  \BibitemOpen
  \bibfield  {author} {\bibinfo {author} {\bibfnamefont {R.}~\bibnamefont
  {Drost}}, \bibinfo {author} {\bibfnamefont {T.}~\bibnamefont {Ojanen}},
  \bibinfo {author} {\bibfnamefont {A.}~\bibnamefont {Harju}}, \ and\ \bibinfo
  {author} {\bibfnamefont {P.}~\bibnamefont {Liljeroth}},\ }\href
  {http://dx.doi.org/10.1038/nphys4080} {\bibfield  {journal} {\bibinfo
  {journal} {Nature Physics}\ }\textbf {\bibinfo {volume} {13}},\ \bibinfo
  {pages} {668} (\bibinfo {year} {2017})}\BibitemShut {NoStop}%
\bibitem [{\citenamefont {Slot}\ \emph {et~al.}(2017)\citenamefont {Slot},
  \citenamefont {Gardenier}, \citenamefont {Jacobse}, \citenamefont {van
  Miert}, \citenamefont {Kempkes}, \citenamefont {Zevenhuizen}, \citenamefont
  {Smith}, \citenamefont {Vanmaekelbergh},\ and\ \citenamefont
  {Swart}}]{Slot:2017}%
  \BibitemOpen
  \bibfield  {author} {\bibinfo {author} {\bibfnamefont {M.~R.}\ \bibnamefont
  {Slot}}, \bibinfo {author} {\bibfnamefont {T.~S.}\ \bibnamefont {Gardenier}},
  \bibinfo {author} {\bibfnamefont {P.~H.}\ \bibnamefont {Jacobse}}, \bibinfo
  {author} {\bibfnamefont {G.~C.~P.}\ \bibnamefont {van Miert}}, \bibinfo
  {author} {\bibfnamefont {S.~N.}\ \bibnamefont {Kempkes}}, \bibinfo {author}
  {\bibfnamefont {S.~J.~M.}\ \bibnamefont {Zevenhuizen}}, \bibinfo {author}
  {\bibfnamefont {C.~M.}\ \bibnamefont {Smith}}, \bibinfo {author}
  {\bibfnamefont {D.}~\bibnamefont {Vanmaekelbergh}}, \ and\ \bibinfo {author}
  {\bibfnamefont {I.}~\bibnamefont {Swart}},\ }\href
  {http://dx.doi.org/10.1038/nphys4105} {\bibfield  {journal} {\bibinfo
  {journal} {Nature Physics}\ }\textbf {\bibinfo {volume} {13}},\ \bibinfo
  {pages} {672} (\bibinfo {year} {2017})}\BibitemShut {NoStop}%
\bibitem [{\citenamefont {Asteria}\ \emph {et~al.}()\citenamefont {Asteria},
  \citenamefont {Tran}, \citenamefont {Tarnowski}, \citenamefont {Rem},
  \citenamefont {Fl\"aschner}, \citenamefont {Sengstock}, \citenamefont
  {Goldman},\ and\ \citenamefont {Weitenberg}}]{Asteria:2018}%
  \BibitemOpen
  \bibfield  {author} {\bibinfo {author} {\bibfnamefont {L.}~\bibnamefont
  {Asteria}}, \bibinfo {author} {\bibfnamefont {O.~T.}\ \bibnamefont {Tran},
  \bibfnamefont {T.D.}}, \bibinfo {author} {\bibfnamefont {M.}~\bibnamefont
  {Tarnowski}}, \bibinfo {author} {\bibfnamefont {B.}~\bibnamefont {Rem}},
  \bibinfo {author} {\bibfnamefont {N.}~\bibnamefont {Fl\"aschner}}, \bibinfo
  {author} {\bibfnamefont {K.}~\bibnamefont {Sengstock}}, \bibinfo {author}
  {\bibfnamefont {N.}~\bibnamefont {Goldman}}, \ and\ \bibinfo {author}
  {\bibfnamefont {C.}~\bibnamefont {Weitenberg}},\ }\href@noop {} {\ }\Eprint
  {http://arxiv.org/abs/1805.11077} {arXiv:1805.11077} \BibitemShut {NoStop}%
\end{thebibliography}%


\begin{thebibliography}{5}%
\makeatletter
\providecommand \@ifxundefined [1]{%
 \@ifx{#1\undefined}
}%
\providecommand \@ifnum [1]{%
 \ifnum #1\expandafter \@firstoftwo
 \else \expandafter \@secondoftwo
 \fi
}%
\providecommand \@ifx [1]{%
 \ifx #1\expandafter \@firstoftwo
 \else \expandafter \@secondoftwo
 \fi
}%
\providecommand \natexlab [1]{#1}%
\providecommand \enquote  [1]{``#1''}%
\providecommand \bibnamefont  [1]{#1}%
\providecommand \bibfnamefont [1]{#1}%
\providecommand \citenamefont [1]{#1}%
\providecommand \href@noop [0]{\@secondoftwo}%
\providecommand \href [0]{\begingroup \@sanitize@url \@href}%
\providecommand \@href[1]{\@@startlink{#1}\@@href}%
\providecommand \@@href[1]{\endgroup#1\@@endlink}%
\providecommand \@sanitize@url [0]{\catcode `\\12\catcode `\$12\catcode
  `\&12\catcode `\#12\catcode `\^12\catcode `\_12\catcode `\%12\relax}%
\providecommand \@@startlink[1]{}%
\providecommand \@@endlink[0]{}%
\providecommand \url  [0]{\begingroup\@sanitize@url \@url }%
\providecommand \@url [1]{\endgroup\@href {#1}{\urlprefix }}%
\providecommand \urlprefix  [0]{URL }%
\providecommand \Eprint [0]{\href }%
\providecommand \doibase [0]{http://dx.doi.org/}%
\providecommand \selectlanguage [0]{\@gobble}%
\providecommand \bibinfo  [0]{\@secondoftwo}%
\providecommand \bibfield  [0]{\@secondoftwo}%
\providecommand \translation [1]{[#1]}%
\providecommand \BibitemOpen [0]{}%
\providecommand \bibitemStop [0]{}%
\providecommand \bibitemNoStop [0]{.\EOS\space}%
\providecommand \EOS [0]{\spacefactor3000\relax}%
\providecommand \BibitemShut  [1]{\csname bibitem#1\endcsname}%
\let\auto@bib@innerbib\@empty
\bibitem [{\citenamefont {Peotta}\ and\ \citenamefont
  {T\"orm\"a}(2015)}]{Peotta:2015}%
  \BibitemOpen
  \bibfield  {author} {\bibinfo {author} {\bibfnamefont {S.}~\bibnamefont
  {Peotta}}\ and\ \bibinfo {author} {\bibfnamefont {P.}~\bibnamefont
  {T\"orm\"a}},\ }\href {http://dx.doi.org/10.1038/ncomms9944} {\bibfield
  {journal} {\bibinfo  {journal} {Nat. Commun.}\ }\textbf {\bibinfo {volume}
  {6}},\ \bibinfo {pages} {8944} (\bibinfo {year} {2015})}\BibitemShut
  {NoStop}%
\bibitem [{\citenamefont {Tovmasyan}\ \emph {et~al.}(2016)\citenamefont
  {Tovmasyan}, \citenamefont {Peotta}, \citenamefont {T\"orm\"a},\ and\
  \citenamefont {Huber}}]{Tovmasyan:2016}%
  \BibitemOpen
  \bibfield  {author} {\bibinfo {author} {\bibfnamefont {M.}~\bibnamefont
  {Tovmasyan}}, \bibinfo {author} {\bibfnamefont {S.}~\bibnamefont {Peotta}},
  \bibinfo {author} {\bibfnamefont {P.}~\bibnamefont {T\"orm\"a}}, \ and\
  \bibinfo {author} {\bibfnamefont {S.~D.}\ \bibnamefont {Huber}},\ }\href
  {\doibase 10.1103/PhysRevB.94.245149} {\bibfield  {journal} {\bibinfo
  {journal} {Phys. Rev. B}\ }\textbf {\bibinfo {volume} {94}},\ \bibinfo
  {pages} {245149} (\bibinfo {year} {2016})}\BibitemShut {NoStop}%
\bibitem [{\citenamefont {Liang}\ \emph {et~al.}(2017)\citenamefont {Liang},
  \citenamefont {Vanhala}, \citenamefont {Peotta}, \citenamefont {Siro},
  \citenamefont {Harju},\ and\ \citenamefont {T\"orm\"a}}]{Liang:2017a}%
  \BibitemOpen
  \bibfield  {author} {\bibinfo {author} {\bibfnamefont {L.}~\bibnamefont
  {Liang}}, \bibinfo {author} {\bibfnamefont {T.~I.}\ \bibnamefont {Vanhala}},
  \bibinfo {author} {\bibfnamefont {S.}~\bibnamefont {Peotta}}, \bibinfo
  {author} {\bibfnamefont {T.}~\bibnamefont {Siro}}, \bibinfo {author}
  {\bibfnamefont {A.}~\bibnamefont {Harju}}, \ and\ \bibinfo {author}
  {\bibfnamefont {P.}~\bibnamefont {T\"orm\"a}},\ }\href {\doibase
  10.1103/PhysRevB.95.024515} {\bibfield  {journal} {\bibinfo  {journal} {Phys.
  Rev. B}\ }\textbf {\bibinfo {volume} {95}},\ \bibinfo {pages} {024515}
  (\bibinfo {year} {2017})}\BibitemShut {NoStop}%
\bibitem [{\citenamefont {Aidelsburger}\ \emph {et~al.}(2013)\citenamefont
  {Aidelsburger}, \citenamefont {Atala}, \citenamefont {Lohse}, \citenamefont
  {Barreiro}, \citenamefont {Paredes},\ and\ \citenamefont
  {Bloch}}]{HarperModelBloch}%
  \BibitemOpen
  \bibfield  {author} {\bibinfo {author} {\bibfnamefont {M.}~\bibnamefont
  {Aidelsburger}}, \bibinfo {author} {\bibfnamefont {M.}~\bibnamefont {Atala}},
  \bibinfo {author} {\bibfnamefont {M.}~\bibnamefont {Lohse}}, \bibinfo
  {author} {\bibfnamefont {J.~T.}\ \bibnamefont {Barreiro}}, \bibinfo {author}
  {\bibfnamefont {B.}~\bibnamefont {Paredes}}, \ and\ \bibinfo {author}
  {\bibfnamefont {I.}~\bibnamefont {Bloch}},\ }\href {\doibase
  10.1103/PhysRevLett.111.185301} {\bibfield  {journal} {\bibinfo  {journal}
  {Phys. Rev. Lett.}\ }\textbf {\bibinfo {volume} {111}},\ \bibinfo {pages}
  {185301} (\bibinfo {year} {2013})}\BibitemShut {NoStop}%
\bibitem [{\citenamefont {Miyake}\ \emph {et~al.}(2013)\citenamefont {Miyake},
  \citenamefont {Siviloglou}, \citenamefont {Kennedy}, \citenamefont {Burton},\
  and\ \citenamefont {Ketterle}}]{HarperModelKetterle}%
  \BibitemOpen
  \bibfield  {author} {\bibinfo {author} {\bibfnamefont {H.}~\bibnamefont
  {Miyake}}, \bibinfo {author} {\bibfnamefont {G.~A.}\ \bibnamefont
  {Siviloglou}}, \bibinfo {author} {\bibfnamefont {C.~J.}\ \bibnamefont
  {Kennedy}}, \bibinfo {author} {\bibfnamefont {W.~C.}\ \bibnamefont {Burton}},
  \ and\ \bibinfo {author} {\bibfnamefont {W.}~\bibnamefont {Ketterle}},\
  }\href {\doibase 10.1103/PhysRevLett.111.185302} {\bibfield  {journal}
  {\bibinfo  {journal} {Phys. Rev. Lett.}\ }\textbf {\bibinfo {volume} {111}},\
  \bibinfo {pages} {185302} (\bibinfo {year} {2013})}\BibitemShut {NoStop}%
\end{thebibliography}%

\end{document}


\title{Quantum metric and effective mass of a two-body bound state in a flat band\\ Supplemental Material}


\author{P. T{\"o}rm{\"a}}
\email[]{paivi.torma@aalto.fi}
\author{L. Liang}
\email[]{long.liang@aalto.fi}
\author{S. Peotta}
\affiliation{Department of Applied Physics, Aalto University, FI-00076 Aalto, Finland}


\newcommand{\PT}[1]{\textcolor{red}{{\bf PT: #1 }}}

\date{\today}



\pacs{}
\maketitle

\section{Derivation based on solving the Schr\"odinger equation using a separable potential}

We start from the Schr\"odinger equation
\begin{equation}
\left[ T_1 + T_2 + \lambda V(1,2) \right] |\psi(1,2)\rangle = E |\psi(1,2)\rangle ,  \label{OrigSchrode}
\end{equation}
where $T_1 + T_2$ contains the kinetic energies of the two particles (fermions or bosons) and the periodic potential, and $\lambda V(1,2)$
is the interaction potential between the particles.
The solution without the interaction potential is the two-particle state given by
$H_0 |\varphi_n\rangle = (T_1 + T_2) |\varphi_n\rangle = E_n  |\varphi_n\rangle$,
where $n$ contains all quantum numbers (band index, lattice momenta, spin) of the two-particle state. 

Let $|\varphi_0(1,2)\rangle$ denote the state of the particles in absence of interactions and $E_0$ the corresponding energy. We consider a flat band where all other states with different quantum numbers (but the same band index) have the same energy, $E_n=E_0$ (or $E_n \simeq E_0$). We denote by $n$ the states in this flat band and by $n'$ those in other dispersive or flat bands. The solution that fulfills the Schr\"odinger equation (\ref{OrigSchrode}) is given by
\begin{align}
|\psi (1,2)\rangle =& |\varphi_0(1,2)\rangle + \sum_{n\neq 0} \frac{|\varphi_n\rangle}{E-E_0} \langle \varphi_n | \lambda V| \psi(1,2)\rangle \nonumber \\
+& \sum_{n'} \frac{|\varphi_{n'}\rangle}{E-E_{n'}} \langle \varphi_{n'} | \lambda V| \psi(1,2)\rangle   \label{multibandSchrode} \\
E-E_0 =& \langle \varphi_0 | \lambda V| \psi(1,2)\rangle .  \label{initialEnergy}
\end{align}

The proof assumes that the eigenfunctions are orthogonal and complete, allowing to write (labels 1, 2 are dropped here)
\begin{align}
(H_0 - E) |\psi\rangle =& (E_0 - E) |\varphi_0\rangle + \sum_{n\neq 0} \frac{(E_n-E)|\varphi_n\rangle}{E-E_n} \langle \varphi_n | \lambda V| \psi \rangle   \\
=& (E_0 - E) |\varphi_0\rangle + |\varphi_0\rangle \langle \varphi_0 | \lambda V| \psi\rangle - \sum_{n = 0}^{n_{max}} \frac{(E-E_n)|\varphi_n\rangle}{E-E_n} \langle \varphi_n | \lambda V| \psi\rangle .
\end{align}
Now one can use $\sum_{n = 0}^{n_{max}} | \varphi_n \rangle \langle \varphi_n | = 1 $ to simplify the last term so that
\begin{align}
(H_0 - E) |\psi\rangle 
=& (E_0 - E) |\varphi_0\rangle + |\varphi_0\rangle \langle \varphi_0 | \lambda V| \psi\rangle -   \lambda V| \psi\rangle \\
\Rightarrow (H_0 + \lambda V) |\psi\rangle =& E |\psi\rangle + (E_0 - E +  \langle \varphi_0 | \lambda V| \psi\rangle) |\varphi_0\rangle ,
\end{align}
which gives (\ref{OrigSchrode}), with the last term zero by (\ref{initialEnergy}). Due to the use of the completeness relation, it is essential to include all bands, the eigenstates of a single band in a multiband system are not complete. 

We assume the isolated flat band limit: The lowest/highest energies $E'_{min/max}$ of the bands above/below the flat band are separated from it by a band gap that is larger than the interactions, $|E_0-E'_{min/max}|>>|\lambda|$. If we further assume that $E$ is close to $E_0$, which should be the case for weak interactions, then $|E-E'_{min/max}|>>|\lambda|$ and the last term of (\ref{multibandSchrode}) becomes negligible. The validity of this approximation can be confirmed by checking that for the final result indeed $|E-E_0|<<|E-E'_{min/max}|$. Within the isolated flat band approximation we proceed with
\begin{align}
|\psi (1,2)\rangle =& |\varphi_0(1,2)\rangle + \frac{1}{E-E_0}\sum_{n\neq 0} |\varphi_n\rangle \langle \varphi_n | \lambda V| \psi(1,2)\rangle \label{Start1} \\
E-E_0 =& \langle \varphi_0 | \lambda V| \psi(1,2)\rangle , \label{Start2}
\end{align}
where $n$ refers to quantum numbers in the isolated flat band.

We use the Bloch functions $e^{i \mathbf{k} \cdot \mathbf{x}} m_\mathbf{k}(\mathbf{x})$ of the flat band where $\mathbf{k}$ is the lattice momentum and the band and spin indices are not marked explicitly. Then
\begin{align}
\varphi_0(\mathbf{x}_1,\mathbf{x}_2) = \langle \mathbf{x}_1,\mathbf{x}_2 |\varphi_0(1,2)\rangle =& e^{i \mathbf{k}_1 \cdot \mathbf{x}_1} m_{\mathbf{k}_1} (\mathbf{x}_1) 
e^{i \mathbf{k}_2 \cdot \mathbf{x}_2} m_{\mathbf{k}_2} (\mathbf{x}_2) \nonumber  \\
=& e^{i \mathbf{q} \cdot \mathbf{R}} e^{i \mathbf{k} \cdot \mathbf{r}} m_{\mathbf{k}+\frac{\mathbf{q}}{2}} (\mathbf{x}_1)  m_{-\mathbf{k}+\frac{\mathbf{q}}{2}} (\mathbf{x}_2) . 
\end{align}
The plane wave part of the wavefunction can be written in terms of the center-of-mass (COM) and relative coordinates $\mathbf{q}= \mathbf{k_1} + \mathbf{k_2}$, $\mathbf{k}= (\mathbf{k_1} - \mathbf{k_2})/2$, $\mathbf{R}= (\mathbf{x_1} + \mathbf{x_2})/2$, $\mathbf{r}= \mathbf{x_1} - \mathbf{x_2}$. The Bloch functions $m_{\mathbf{k}} (\mathbf{x})$ are assumed normalized by the system volume, and the volume does not appear explicitly in the calculation in this section. 
We then consider the case where particle one has spin up and the particle two spin down, and use  $m^\uparrow_{\mathbf{k}}= m^{\downarrow *}_{-\mathbf{k}}\equiv m_{\mathbf{k}}$ which is consequence of time reversal symmetry. This brings the above formula into
\begin{align}
\varphi_0(\mathbf{x}_1,\mathbf{x}_2) 
=& e^{i \mathbf{q} \cdot \mathbf{R}} e^{i \mathbf{k} \cdot \mathbf{r}} m_{\mathbf{k}+\frac{\mathbf{q}}{2}} (\mathbf{x}_1)  m^*_{\mathbf{k}-\frac{\mathbf{q}}{2}} (\mathbf{x}_2) . 
\end{align}

We introduce a definition of the total wavefunction:
\begin{align}
\psi (\mathbf{x}_1,\mathbf{x}_2) = e^{i \mathbf{q} \cdot \mathbf{R}} \psi_{\mathbf{q},\mathbf{k}} (\mathbf{x}_1, \mathbf{x}_2) .
\end{align}
Since in a lattice we cannot assume a separation of the COM and relative coordinates, this has to be understood as just a redefinition (multiplication of wave function by a phase factor). 
The interaction potential $V$ is assumed to have the same periodicity in its COM coordinate dependence as the lattice (this allows a large class of possible potentials, also orbital-dependent ones). Then the total lattice momentum $\mathbf{q}= \mathbf{k_1} + \mathbf{k_2}$ is conserved. 
 
The equations (\ref{Start1})-(\ref{Start2}) are now
\begin{align}
e^{i \mathbf{q} \cdot \mathbf{R}} \psi_{\mathbf{q}, \mathbf{k}} (\mathbf{x}_1,\mathbf{x}_2) =& e^{i \mathbf{q} \cdot \mathbf{R}} e^{i \mathbf{k} \cdot \mathbf{r}} m_{\mathbf{k}+\frac{\mathbf{q}}{2}} (\mathbf{x}_1)  m^*_{\mathbf{k}-\frac{\mathbf{q}}{2}} (\mathbf{x}_2) \\
+& \frac{\lambda}{E-E_0} \sum_{\mathbf{t} \neq \mathbf{k}} e^{i \mathbf{q} \cdot \mathbf{R}} e^{i \mathbf{t} \cdot \mathbf{r}} 
m_{\mathbf{t}+\frac{\mathbf{q}}{2}} (\mathbf{x}_1)  m^*_{\mathbf{t}-\frac{\mathbf{q}}{2}} (\mathbf{x}_2) \\
\times & \langle 
e^{i \mathbf{q} \cdot \mathbf{R}} e^{i \mathbf{t} \cdot \mathbf{r}} m_{\mathbf{t}+\frac{\mathbf{q}}{2}} (\mathbf{x}_1)  m^*_{\mathbf{t}-\frac{\mathbf{q}}{2}} (\mathbf{x}_2)  | V | e^{i \mathbf{q} \cdot \mathbf{R}} \psi_{\mathbf{q}, \mathbf{k}} (\mathbf{x}_1,\mathbf{x}_2)\rangle \\
E-E_0 =& \lambda \langle e^{i \mathbf{q} \cdot \mathbf{R}} e^{i \mathbf{k} \cdot \mathbf{r}} 
 m_{\mathbf{k}+\frac{\mathbf{q}}{2}} (\mathbf{x}_1)  m^*_{\mathbf{k}-\frac{\mathbf{q}}{2}} (\mathbf{x}_2) | V| e^{i \mathbf{q} \cdot \mathbf{R}} \psi_{\mathbf{q}, \mathbf{k}} (\mathbf{x}_1,\mathbf{x}_2) \rangle  .
\end{align}
It thus follows that the COM momentum $\mathbf{q}$ dependence is only left in the periodic part of the Bloch function and in $\psi_{\mathbf{q}, \mathbf{k}} (\mathbf{x}_1,\mathbf{x}_2)$:
\begin{align}
\psi_{\mathbf{q}, \mathbf{k}} (\mathbf{x}_1,\mathbf{x}_2) =& e^{i \mathbf{k} \cdot \mathbf{r}} 
 m_{\mathbf{k}+\frac{\mathbf{q}}{2}} (\mathbf{x}_1)  m^*_{\mathbf{k}-\frac{\mathbf{q}}{2}} (\mathbf{x}_2) \\
+& \frac{\lambda}{E-E_0} 
 \sum_{\mathbf{t}\neq \mathbf{k}} 
 e^{i \mathbf{t} \cdot \mathbf{r}} 
 m_{\mathbf{t}+\frac{\mathbf{q}}{2}} (\mathbf{x}_1)  m^*_{\mathbf{t}-\frac{\mathbf{q}}{2}} (\mathbf{x}_2) \langle 
 e^{i \mathbf{t} \cdot \mathbf{r}} 
 m_{\mathbf{t}+\frac{\mathbf{q}}{2}} (\mathbf{x}_1)  m^*_{\mathbf{t}-\frac{\mathbf{q}}{2}} (\mathbf{x}_2) | V |  \psi_{\mathbf{q}, \mathbf{k}} (\mathbf{x}_1,\mathbf{x}_2)\rangle \\
E-E_0 =& \lambda \langle  e^{i \mathbf{k} \cdot \mathbf{r}} 
 m_{\mathbf{k}+\frac{\mathbf{q}}{2}} (\mathbf{x}_1)  m^*_{\mathbf{k}-\frac{\mathbf{q}}{2}} (\mathbf{x}_2) | V|  \psi_{\mathbf{q}, \mathbf{k}} (\mathbf{x}_1,\mathbf{x}_2)\rangle  .  \label{EwithBloch}
\end{align}

We consider a general interaction potential $V=V(\mathbf{x}_1, \mathbf{x}_2)$, and make the separable potential approximation, namely $V(\mathbf{x}_1, \mathbf{x}_2) \longrightarrow u(\mathbf{x}_1, \mathbf{x}_2)u(\mathbf{x}_1', \mathbf{x}_2')$, and assume $u$ real. Strictly speaking, such separation is justifiable only for a delta function potential (contact interactions), but it may be a good approximation in other cases as well. In the end of the calculation, the original potential is reintroduced by
\begin{equation}
u(\mathbf{x}_1, \mathbf{x}_2)u(\mathbf{x}_1', \mathbf{x}_2') \longrightarrow 
V(\mathbf{x}_1, \mathbf{x}_2) \delta(\mathbf{x}_1-\mathbf{x}_1') \delta(\mathbf{x}_2-\mathbf{x}_2')  .
\end{equation}

The matrix element 
\begin{align}
\int d\textbf{x}_1 d\textbf{x}_2 e^{-i \mathbf{k} \cdot \mathbf{r}}
 m^*_{\mathbf{k}+\frac{\mathbf{q}}{2}} (\mathbf{x}_1)  m_{\mathbf{k}-\frac{\mathbf{q}}{2}} (\mathbf{x}_2) V(\mathbf{x}_1, \mathbf{x}_2)   \psi_{\mathbf{q},\mathbf{k}} (\textbf{x}_1,\textbf{x}_2)  
\end{align}
in the equation (\ref{EwithBloch}) becomes under the separable potential approximation
\begin{align}
\int d\textbf{x}_1 d\textbf{x}_2 d\textbf{x}'_1 d\textbf{x}'_2 e^{-i \mathbf{k} \cdot \mathbf{r}} m^*_{\mathbf{k}+\frac{\mathbf{q}}{2}} (\mathbf{x}_1)  m_{\mathbf{k}-\frac{\mathbf{q}}{2}} (\mathbf{x}_2)   u(\mathbf{x}_1, \mathbf{x}_2)u(\mathbf{x}_1', \mathbf{x}_2') \psi_{\mathbf{q},\mathbf{k}}(\textbf{x}'_1, \textbf{x}'_2)  .
\end{align}

We also define 
\begin{align}
\tilde{u} (\mathbf{k_1},\mathbf{k_2}) = \int d\textbf{x}_1 d\textbf{x}_2 e^{-i \mathbf{k} \cdot \mathbf{r}} m^*_{\mathbf{k}+\frac{\mathbf{q}}{2}} (\mathbf{x}_1)  m_{\mathbf{k}-\frac{\mathbf{q}}{2}} (\mathbf{x}_2) 
u(\mathbf{x}_1, \mathbf{x}_2) . \nonumber
\end{align}

Now we can proceed with
\begin{align}
\psi_{\mathbf{q}, \mathbf{k}} (\mathbf{x}_1,\mathbf{x}_2) =&  e^{i \mathbf{k} \cdot \mathbf{r}} m_{\mathbf{k}+\frac{\mathbf{q}}{2}} (\mathbf{x}_1)  m^*_{\mathbf{k}-\frac{\mathbf{q}}{2}} (\mathbf{x}_2) \\
+& \frac{\lambda}{E-E_0} \sum_{\mathbf{t} \neq \mathbf{k}} e^{i \mathbf{t} \cdot \mathbf{r}} 
m_{\mathbf{t}+\frac{\mathbf{q}}{2}} (\mathbf{x}_1)  m^*_{\mathbf{t}-\frac{\mathbf{q}}{2}} (\mathbf{x}_2)  \\
\times & \int 
d\textbf{x}'_1 d\textbf{x}'_2 d\textbf{x}^{''}_1 d\textbf{x}^{''}_2 e^{-i \mathbf{t} \cdot \mathbf{r}'} m^*_{\mathbf{t}+\frac{\mathbf{q}}{2}} (\mathbf{x}'_1)  m_{\mathbf{t}-\frac{\mathbf{q}}{2}} (\mathbf{x}'_2)  u(\textbf{x}'_1,\textbf{x}'_2) 
u(\textbf{x}^{''}_1,\textbf{x}^{''}_2) \psi_{\mathbf{q},\mathbf{k}}(\textbf{x}^{''}_1, \textbf{x}^{''}_2) .
\end{align}
We obtain
\begin{align}
E-E_0 =& \lambda \int d\textbf{x}_1 d\textbf{x}_2 d\textbf{x}'_1 d\textbf{x}'_2 e^{-i \mathbf{k} \cdot \mathbf{r}} m^*_{\mathbf{k}+\frac{\mathbf{q}}{2}} (\mathbf{x}_1)  m_{\mathbf{k}-\frac{\mathbf{q}}{2}} (\mathbf{x}_2)  u(\textbf{x}_1,\textbf{x}_2) 
u(\textbf{x}'_1,\textbf{x}'_2) \psi_{\mathbf{q},\mathbf{k}}(\textbf{x}'_1, \textbf{x}'_2) \\
=& \lambda \tilde{u}(\mathbf{q},\mathbf{k}) \int d\textbf{x}_1 d\textbf{x}_2 
u(\textbf{x}_1,\textbf{x}_2)  \psi_{\mathbf{q},\mathbf{k}}(\textbf{x}_1, \textbf{x}_2) \label{EnEq} \\
=&  \lambda \tilde{u}(\mathbf{q},\mathbf{k}) \left( \int d\textbf{x}_1 d\textbf{x}_2 
u(\textbf{x}_1,\textbf{x}_2) e^{i \mathbf{k} \cdot \mathbf{r}} m_{\mathbf{k}+\frac{\mathbf{q}}{2}} (\mathbf{x}_1)  m^*_{\mathbf{k}-\frac{\mathbf{q}}{2}} (\mathbf{x}_2) \right. \\
+& \frac{\lambda}{E-E_0} \int d\textbf{x}_1 d\textbf{x}_2 
u(\textbf{x}_1,\textbf{x}_2)  \sum_{\mathbf{t}\neq \mathbf{k}} 
e^{i \mathbf{t} \cdot \mathbf{r}} 
m_{\mathbf{t}+\frac{\mathbf{q}}{2}} (\mathbf{x}_1)  m^*_{\mathbf{t}-\frac{\mathbf{q}}{2}} (\mathbf{x}_2) \\
\times & \left. \int 
d\textbf{x}'_1 d\textbf{x}'_2 d\textbf{x}^{''}_1 d\textbf{x}^{''}_2 e^{-i \mathbf{t} \cdot \mathbf{r}'} m^*_{\mathbf{t}+\frac{\mathbf{q}}{2}} (\mathbf{x}'_1)  m_{\mathbf{t}-\frac{\mathbf{q}}{2}} (\mathbf{x}'_2)  u(\textbf{x}'_1,\textbf{x}'_2)u(\textbf{x}^{''}_1,\textbf{x}^{''}_2) \psi_{\mathbf{q},\mathbf{k}} (\textbf{x}^{''}_1, \textbf{x}^{''}_2) \right) \\
=& \lambda |\tilde{u}(\mathbf{q},\mathbf{k})|^2 + \frac{\lambda^2}{E-E_0} 
\sum_{\mathbf{t}\neq \mathbf{k}} |\tilde{u}(\mathbf{q},\mathbf{t})|^2 \tilde{u}(\mathbf{q},\mathbf{k}) \int d\textbf{x}_1 d\textbf{x}_2  u(\textbf{x}_1,\textbf{x}_2) \psi_{\mathbf{q},\mathbf{k}} (\textbf{x}_1, \textbf{x}_2) \\
=& \lambda |\tilde{u}(\mathbf{q},\mathbf{k})|^2 + \frac{\lambda^2}{E-E_0} 
\sum_{\mathbf{t} \neq \mathbf{k}} |\tilde{u}(\mathbf{q},\mathbf{t})|^2 \frac{E-E_0}{\lambda} .
\end{align}
In the last step, equation (\ref{EnEq}) was used. This gives
\begin{equation}
E-E_0 = \lambda \sum_{\mathbf{k}} |\tilde{u} (\mathbf{q},\mathbf{k})|^2   . \label{ApprResult}
\end{equation}
In $|\tilde{u} (\mathbf{q},\mathbf{k})|^2$ we can replace the separable potential by the original one to obtain
\begin{align}
 &|\tilde{u} (\mathbf{q},\mathbf{k})|^2 = \nonumber \\   
     &\int d\textbf{x}_1 d\textbf{x}_2d\textbf{x}'_1 d\textbf{x}'_2 e^{-i \mathbf{k} \cdot (\mathbf{r}-\mathbf{r}')} m^*_{\mathbf{k}+\frac{\mathbf{q}}{2}} (\mathbf{x}_1)  m_{\mathbf{k}-\frac{\mathbf{q}}{2}} (\mathbf{x}_2) m_{\mathbf{k}+\frac{\mathbf{q}}{2}} (\mathbf{x}'_1)  m^*_{\mathbf{k}-\frac{\mathbf{q}}{2}} (\mathbf{x}'_2)
u(\mathbf{x}_1, \mathbf{x}_2) u(\mathbf{x}'_1, \mathbf{x}'_2) \nonumber \\
&\longrightarrow \int d\textbf{x}_1 d\textbf{x}_2d\textbf{x}'_1 d\textbf{x}'_2 e^{-i \mathbf{k} \cdot (\mathbf{r}-\mathbf{r}')} m^*_{\mathbf{k}+\frac{\mathbf{q}}{2}} (\mathbf{x}_1)  m_{\mathbf{k}-\frac{\mathbf{q}}{2}} (\mathbf{x}_2) m_{\mathbf{k}+\frac{\mathbf{q}}{2}} (\mathbf{x}'_1)  m^*_{\mathbf{k}-\frac{\mathbf{q}}{2}} (\mathbf{x}'_2)
V(\mathbf{x}_1,\mathbf{x}_2)\delta (\mathbf{x}_1-\mathbf{x}'_1) \delta (\mathbf{x}_2-\mathbf{x}'_2) \nonumber \\
&= \int d\textbf{x}_1 d\textbf{x}_2  m^*_{\mathbf{k}+\frac{\mathbf{q}}{2}} (\mathbf{x}_1)  m^*_{\mathbf{k}-\frac{\mathbf{q}}{2}} (\mathbf{x}_2) m_{\mathbf{k}+\frac{\mathbf{q}}{2}} (\mathbf{x}_1)  m_{\mathbf{k}-\frac{\mathbf{q}}{2}} (\mathbf{x}_2)
V(\mathbf{x}_1,\mathbf{x}_2) .
\end{align}

We continue to process the result further by assuming that the interaction is of contact potential type, namely $V(\mathbf{x}_1, \mathbf{x}_2)=\delta(\mathbf{x}_1 - \mathbf{x}_2) V(\mathbf{x}_1)$. This still keeps the possibility that the interaction depends on position, that is, the strength of the contact interaction may be orbital dependent. The energy becomes 
\begin{equation}
E-E_0 =  \lambda  \sum_{\mathbf{k}} \int d\mathbf{x} V(\mathbf{x}) | m_{\mathbf{k} + \frac{\mathbf{q}}{2}} (\mathbf{x}) m_{\mathbf{k} - \frac{\mathbf{q}}{2}} (\mathbf{x})|^2 . \label{NonUnifFinal}
\end{equation}

To inspect the possibility of a finite effective mass, we make an expansion in the pair momentum, that is, a small $q$ Taylor series, to the Bloch function: $m_{\mathbf{k} \pm \frac{\mathbf{q}}{2}} = m_{\mathbf{k}} \pm \frac{q_i}{2} \partial_i m_{\mathbf{k}}  + \frac{1}{8} q_i q_j \partial_i \partial_j m_{\mathbf{k}}  + \mathcal{O}(q^3)$. Summation is assumed over repeated indices and $\partial_i \equiv \partial/\partial k_i$. This gives
\begin{align}
E-E_0 &= \lambda  \sum_{\mathbf{k}} \int d\mathbf{x} V(\mathbf{x}) 
\left[ |m_{\mathbf{k}} (\mathbf{x})|^4 \right. \nonumber \\
&+ \frac{q_i q_j}{4} \left. \left( |m_{\mathbf{k}} (\mathbf{x})|^2 m_{\mathbf{k}} (\mathbf{x}) \partial_i \partial_j m^*_{\mathbf{k}} (\mathbf{x}) - (m_{\mathbf{k}} (\mathbf{x}))^2 \partial_i m^*_{\mathbf{k}} (\mathbf{x}) \partial_j m^*_{\mathbf{k}} (\mathbf{x}) + h.c. 
 \right)\right] \nonumber \\
&= \lambda \sum_{\mathbf{k}} \int d\mathbf{x} V(\mathbf{x}) 
\left[ P_\mathbf{k}(\mathbf{x})^2   -  \frac{q_i q_j}{4}  \left( \partial_i P_\mathbf{k}(\mathbf{x}) \partial_j P_\mathbf{k}(\mathbf{x}) - P_\mathbf{k}(\mathbf{x}) \partial_i \partial_j P_\mathbf{k}(\mathbf{x})  
 \right)\right]  ,  \label{FinalEnergy}
\end{align}
where $P_\mathbf{k}(\mathbf{x}) = m^*_{\mathbf{k}} (\mathbf{x}) m_{\mathbf{k}} (\mathbf{x}) = \langle \mathbf{x}|m_\mathbf{k}\rangle \langle m_\mathbf{k}|\mathbf{x}\rangle$ is the diagonal element of the projector $P_\mathbf{k}=|m_\mathbf{k}\rangle \langle m_\mathbf{k}| $ to the flat band.
The effective mass tensor is therefore
\begin{align}
    \left[\frac{1}{m^*}\right]_{ij} = - \frac{\lambda}{2} \sum_{\mathbf{k}} &\int d\mathbf{x} V(\mathbf{x}) 
\left(\partial_i P_\mathbf{k}(\mathbf{x}) \partial_j P_\mathbf{k}(\mathbf{x})  - P_\mathbf{k}(\mathbf{x}) \partial_i \partial_j P_\mathbf{k}(\mathbf{x}) \right) .  \label{effMass}
\end{align}
The summation of $\mathbf{k}$ over the first Brillouin zone can be changed to integration, and partial integration can be applied to remove the second derivative. Note that the projectors $P$ are continuous over the whole Brillouin zone even when there would be discontinous jumps in the phases of the Bloch functions (associated with finite Chern numbers, for instance). Therefore this partial integration step is valid in the general case. The result is
\begin{align}
E-E_0 
= \lambda \sum_{\mathbf{k}}  \int d\mathbf{x} V(\mathbf{x}) 
\left( P_\mathbf{k}(\mathbf{x})^2
-   \frac{q_i q_j}{2} 
\partial_i P_\mathbf{k}(\mathbf{x}) \partial_j P_\mathbf{k}(\mathbf{x})   \right) .
\end{align} 
Thus
\begin{align}
    \left[\frac{1}{m^*}\right]_{ij} = - \lambda \sum_{\mathbf{k}} &\int d\mathbf{x} V(\mathbf{x}) 
\partial_i P_\mathbf{k}(\mathbf{x}) \partial_j P_\mathbf{k}(\mathbf{x}) .  \label{effMasssimple}
\end{align}
For $i=j$ this is simply
\begin{align}
    \left[\frac{1}{m^*}\right]_{ii} = -  \lambda \sum_{\mathbf{k}} &\int d\mathbf{x} V(\mathbf{x}) 
\left[\partial_i P_\mathbf{k}(\mathbf{x})\right]^2 .  \label{effMasssimple}
\end{align}
This means that a finite, positive effective mass exists whenever the first derivative of the projector is non-zero at least on one orbital where $V(\mathbf{x})$ is non-zero.

\section{Derivation based on variational ansatz}

In this section, we use the variational method to calculate the two-body energy in a multiband lattice system. Since the two-body potential is translationally invariant or has the same periodicity as the lattice, the total momentum $\mathbf{q}$ of the two particles is conserved. In the isolated band limit,  we can  construct the following variational state
\begin{eqnarray}\label{Eq:WF_variational1}
|\psi_{\mathbf{q}}\rangle=\sum_{\mathbf{k}}A_{\mathbf{k}}c^\dag_{\mathbf{k+q}/2,\uparrow,m} c^\dag_{\mathbf{-k+q}/2,\downarrow,m} |0\rangle.
\end{eqnarray}
Here $c^\dag_{\mathbf{k},\sigma,m}$ creates a Bloch state with momentum $\mathbf{k}$ and spin $\sigma$ in the $m$-th band which is the band we are interested in. 
If the interaction is not  weak compared to the band gap, we can use variational states that mix Bloch bands, and  if all the bands are included in the ansatz, the variational solutions become exact. The isolated flat band approximation is therefore done here by the variational ansatz.

We  construct the functional 
\begin{eqnarray}
F[A_\mathbf{k}]&=&\langle \psi_{\mathbf{q}}|H-E_\mathbf{q}|\psi_{\mathbf{q}} \rangle.
\end{eqnarray}
The  coefficient $A_\mathbf{k}$ and energy $E_\mathbf{q}$ are determined by calculating functional derivative of $F[A_\mathbf{k}]$ with respect to $A_\mathbf{k}$. 
In the second quantized formalism, the two-body potential can be written as 
\begin{eqnarray}
V=\lambda \int \mathrm{d}\mathbf{x}_1\mathrm{d}\mathbf{x}_2\Psi^\dag_{\uparrow}(\mathbf{x}_1)
\Psi^\dag_{\downarrow}(\mathbf{x}_2)V(\mathbf{x}_1-\mathbf{x}_2) \Psi_{\downarrow}(\mathbf{x}_2) \Psi_{\uparrow}(\mathbf{x}_1),
\end{eqnarray}
and the field operator can be expanded by using Bloch functions 
\begin{eqnarray}
\Psi_{\sigma}(\mathbf{x})=\frac{1}{\sqrt{N}_c}\sum_n\sum_{\mathbf{k}} e^{i\mathbf{k}\cdot\mathbf{x}}n_{\mathbf{k},\sigma}(\mathbf{x})c_{\mathbf{k},\sigma,n},
\end{eqnarray}
with $N_c$ being the number of unit cells. Note that the Bloch functions for spin-up and spin-down particles are not necessarily the same. Here $n$ is the band index and $n_{\mathbf{k},\sigma}$ is the periodic part of the Bloch function of the $n$th band. 

We find that (here $m_{\mathbf{k},\sigma}$ is the periodic part of the Bloch function of the band of interest defined by the ansatz, with the band index $m$)
\begin{eqnarray}
F[A_\mathbf{k}]&=&\langle \psi_{\mathbf{q}}|H-E_\mathbf{q}|\psi_{\mathbf{q}} \rangle,\\
&=&\sum_{\mathbf{k}}A_{\mathbf{k}}A^\ast_{\mathbf{k}}(\varepsilon_{\mathbf{k+q}/2,\uparrow,m}+\varepsilon_{\mathbf{-k+q}/2,\downarrow,m}-E_\mathbf{q})\nonumber\\
&& +\frac{ \lambda}{N^2_c}\sum_{\mathbf{k}}\sum_{\mathbf{k}'} A_{\mathbf{k}}A^\ast_{\mathbf{k}'} \int~\mathrm{d}\mathbf{x}_1\mathrm{d}\mathbf{x}_2~V(\mathbf{x}_1-\mathbf{x}_2)e^{i(\mathbf{k}-\mathbf{k}')\cdot(\mathbf{x}_1-\mathbf{x}_2)}\nonumber\\
&& ~~~~~~~~m^\ast_{\mathbf{k'+q}/2,\uparrow}(\mathbf{x}_1)m^\ast_{\mathbf{-k'+q}/2,\downarrow}(\mathbf{x}_2)
m_{\mathbf{-k+q}/2,\downarrow}(\mathbf{x}_2)m_{\mathbf{k+q}/2,\uparrow}(\mathbf{x}_1),\\
&=&\sum_{\mathbf{k}}A_{\mathbf{k}}A^\ast_{\mathbf{k}}(\varepsilon_{\mathbf{k+q}/2,\uparrow,m}+\varepsilon_{\mathbf{-k+q}/2,\downarrow,m}-E_\mathbf{q})+\frac{ 1}{N_c}\sum_{\mathbf{k}}\sum_{\mathbf{k}'} A_{\mathbf{k}}A^\ast_{\mathbf{k}'}V_{\mathbf{k},\mathbf{k}'}(\mathbf{q}).\label{Eq:functional}
\end{eqnarray}
For an exactly flat band, the dispersion $\varepsilon_{\mathbf{k},\sigma,m}$ is independent of the momentum. Here we keep the $\mathbf{k}$ dependence such that our result can be applied to  quasiflat bands. To get Eq.~\eqref{Eq:functional}, 
we have  defined the projected potential $V_{\mathbf{k},\mathbf{k}'}(\mathbf{q})$
\begin{eqnarray}
V_{\mathbf{k},\mathbf{k}'}(\mathbf{q})=&&\frac{ \lambda}{N_c}\int~\mathrm{d}\mathbf{x}_1\mathrm{d}\mathbf{x}_2~V(\mathbf{x}_1-\mathbf{x}_2)e^{i(\mathbf{k}-\mathbf{k}')\cdot(\mathbf{x}_1-\mathbf{x}_2)}\nonumber\\
~~~~~&& m^\ast_{\mathbf{k'+q}/2,\uparrow}(\mathbf{x}_1)m^\ast_{\mathbf{-k'+q}/2,\downarrow}(\mathbf{x}_2)
m_{\mathbf{-k+q}/2,\downarrow}(\mathbf{x}_2)m_{\mathbf{k+q}/2,\uparrow}(\mathbf{x}_1).
\end{eqnarray}

From Eq.~\eqref{Eq:functional} we obtain, by setting the functional derivative with respect to $A_{\mathbf{k}}$ to zero,
\begin{eqnarray}
A_{\mathbf{k}}(\varepsilon_{\mathbf{k+q}/2,\uparrow,m}+\varepsilon_{\mathbf{-k+q}/2,\downarrow,m})+\frac{ 1}{N_c}\sum_{\mathbf{k}'} V_{\mathbf{k},\mathbf{k}'}(\mathbf{q}) A_{\mathbf{k}'}=E_\mathbf{q} A_{\mathbf{k}}.\label{Eq:boundstates}
\end{eqnarray}
It is clear that for an exactly flat band ($\varepsilon_{\mathbf{k+q}/2,\uparrow,m}+\varepsilon_{\mathbf{-k+q}/2,\downarrow,m}=\mathrm{constant}$), the momentum dependence of the bound states solely comes from the Bloch functions in $V_{\mathbf{k},\mathbf{k}'}(\mathbf{q})$. 

\subsection{Tight-binding model}

The equation (\ref{Eq:boundstates}) can be studied for a general type of interaction via numerical solutions. However, to proceed with analytical calculations we use the contact interaction, which is actually an accurate description of most ultracold atom systems, and can be a good first order approximation in others. Contact interaction means we take $V(\mathbf{x}_1-\mathbf{x}_2)=V(\mathbf{x}_1)\delta(\mathbf{x}_1-\mathbf{x}_2)$ and assume that $V(\mathbf{x})$ has the same periodicity as the periodic potential experienced by the particles. For continuum ultracold gas systems, it is natural to assume $V(\mathbf{x})=V$ because the scattering length between the particles is independent of the position. The position dependence is introduced because in the tight binding approximation, the on-site (Hubbard-type) interaction can be orbital dependent.  
Using the contact potential,  the projected potential becomes
\begin{eqnarray}
V_{\mathbf{k},\mathbf{k}'}(\mathbf{q})&=& \frac{\lambda}{N_c}\int~\mathrm{d}\mathbf{x}~V(\mathbf{x}) m^\ast_{\mathbf{k'+q}/2,\uparrow}(\mathbf{x})m^\ast_{\mathbf{-k'+q}/2,\downarrow}(\mathbf{x})
m_{\mathbf{-k+q}/2,\downarrow}(\mathbf{x})m_{\mathbf{k+q}/2,\uparrow}(\mathbf{x}),\\
&=& \lambda \int_{\mathrm{u.c}}~\mathrm{d}\mathbf{x}~ V(\mathbf{x}) m^\ast_{\mathbf{k'+q}/2,\uparrow}(\mathbf{x})m^\ast_{\mathbf{-k'+q}/2,\downarrow}(\mathbf{x})
m_{\mathbf{-k+q}/2,\downarrow}(\mathbf{x})m_{\mathbf{k+q}/2,\uparrow}(\mathbf{x}).\label{Eq:effective_potential}
\end{eqnarray}
where $\mathrm{u.c.}$ denotes unit cell. 
For tight binding models, Eq.~\eqref{Eq:effective_potential} becomes
\begin{eqnarray}\label{Eq:effective_potential_tb}
V_{\mathbf{k},\mathbf{k}'}(\mathbf{q})
&=& \lambda \sum_{\alpha} V_\alpha m^\ast_{\mathbf{k'+q}/2,\uparrow,\alpha}m^\ast_{\mathbf{-k'+q}/2,\downarrow,\alpha}
m_{\mathbf{-k+q}/2,\downarrow,\alpha}m_{\mathbf{k+q}/2,\uparrow,\alpha},
\end{eqnarray}
where $\alpha$ labels the orbitals in the unit cell.

Eq.~\eqref{Eq:boundstates} is an eigenvalue problem and can be solved numerically. To get some physical insights, we discuss two approximations which allow us to get analytical results. In the following we focus on the situations with time reversal symmetry, i.e., $m^\ast_{\mathbf{-k},\downarrow}=m_{\mathbf{k},\uparrow}\equiv m_{\mathbf{k}} $. We assume the band is exactly flat and shift the flat band energy to zero ($\varepsilon_{\mathbf{k+q}/2,\uparrow,m}+\varepsilon_{\mathbf{-k+q}/2,\downarrow,m}=\mathrm{constant}= 0$). 

\subsection{Approximation inspired by the BCS mean-field theory}

We now consider the case where the interaction is orbital independent, $V_\alpha=1$.
Then let us remind of the Bardeen-Cooper-Schrieffer (BCS) approximation ($U$ is the on-site interaction strength, $i$ the position index),
 \begin{eqnarray}
 Uc^\dag_{i\uparrow}c^\dag_{i\downarrow}c_{i\downarrow}c_{i\uparrow}\approx \Delta c_{i\downarrow}c_{i\uparrow}+\Delta^\ast c^\dag_{i\uparrow}c^\dag_{i\downarrow}-|\Delta|^2/U . 
 \end{eqnarray}
We can relate the operator $c^\dag_{i\sigma}$ to the Bloch function $m^\ast_{\mathbf{k},\sigma}$, and the order parameter $\Delta=U\langle c^\dag_{i\uparrow} c^\dag_{i\downarrow}\rangle$ to the expectation value $ \langle m^\ast_{\mathbf{k},\uparrow}  m^\ast_{-\mathbf{k},\downarrow} \rangle$; then, inspired by the structure of the BCS approximation, we may approximate Eq.~\eqref{Eq:effective_potential_tb} by 
\begin{eqnarray}
V_{\mathbf{k},\mathbf{k}'}(\mathbf{q})
&=& \lambda \sum_{\alpha} m^\ast_{\mathbf{k'+q}/2,\alpha}m_{\mathbf{k'-q}/2,\alpha}
m^\ast_{\mathbf{k-q}/2,\alpha}m_{\mathbf{k+q}/2,\alpha},\\
&\approx&
\frac{\lambda}{N_{orb}}\sum_{\alpha,\beta} m^\ast_{\mathbf{k'+q}/2,\alpha}m_{\mathbf{k'-q}/2,\alpha}
m^\ast_{\mathbf{k-q}/2,\beta}m_{\mathbf{k+q}/2,\beta},\label{Eq:effective_potential_tb_approx}\\
&=&
\frac{\lambda}{N_{orb}}\langle m_{\mathbf{k'+q}/2}|m_{\mathbf{k'-q}/2}\rangle \langle m_{\mathbf{k-q}/2}|m_{\mathbf{k+q}/2}\rangle.\label{Eq:projected_potential_pprox1}
\end{eqnarray}
Here $N_{orb}$ is the number of orbitals that compose the flat band. The factor $1/N_{orb}$ is introduced because there are $N^2_{orb}$ terms in Eq.~\eqref{Eq:effective_potential_tb_approx}, while in the original potential there are only $N_{orb}$ terms to be summed.
Moreover, $1/N_{orb}$ appears in the BCS self-consistent equation for $\Delta$, see Eq.~\eqref{Eq:BCS_selfconsistentDelta} in Sec. \ref{Sec:BCS}. These arguments support the approximation done in deriving equation (9) of the main text.

The eigenvalue of the approximated potential, Eq.~\eqref{Eq:projected_potential_pprox1}, can be solved explicitly:
\begin{eqnarray}
&&\frac{\lambda}{N_{orb}}\frac{1}{N_c}\langle m_{\mathbf{k-q}/2}|m_{\mathbf{k+q}/2}\rangle\sum_{\mathbf{k}'}\langle m_{\mathbf{k'+q}/2}|m_{\mathbf{k'-q}/2}\rangle  A_{\mathbf{k}'}=E_{\mathbf{q}}A_{\mathbf{k}}\\
&\longrightarrow & \frac{\lambda}{N_{orb}}\frac{1}{N_c}\sum_{\mathbf{k}}|\langle m_{\mathbf{k-q}/2}|m_{\mathbf{k+q}/2}\rangle|^2\sum_{\mathbf{k}'}\langle m_{\mathbf{k'+q}/2}|m_{\mathbf{k'-q}/2}\rangle  A_{\mathbf{k}'}=E_{\mathbf{q}}\sum_{\mathbf{k}}\langle m_{\mathbf{k+q}/2}|m_{\mathbf{k-q}/2}\rangle A_{\mathbf{k}},\nonumber\\
&\longrightarrow & E_{\mathbf{q}}=\frac{\lambda}{N_{orb}}\frac{1}{N_c}\sum_{\mathbf{k}}|\langle m_{\mathbf{k-q}/2,\uparrow}|m_{\mathbf{k+q}/2,\uparrow}\rangle|^2. \label{Eq:eigenenergy_separable_potential}
\end{eqnarray}
Since the energy Eq.~\eqref{Eq:eigenenergy_separable_potential} is just the  trace of the  potential Eq.~\eqref{Eq:projected_potential_pprox1}, all the other eigenvalues are zero.

We consider now the so-called uniform pairing condition, which in the earlier works~\cite{Peotta:2015,Tovmasyan:2016,Liang:2017a} was defined by having $V_\alpha$ same for all orbitals and also demanding
\begin{eqnarray}
\frac{1}{N_c}\sum_{\mathbf{k}}|m_{\mathbf{k},\alpha}|^2=\frac{1}{N_{orb}} .
\end{eqnarray}
In this case we can show that the zero pair-momentum energy has a simple form: 
\begin{eqnarray}
&&\frac{1}{N_c}\sum_{\mathbf{k}'} V_{\mathbf{k},\mathbf{k}'}(\mathbf{q=0}) A_{\mathbf{k}'}=E_\mathbf{q=0} A_{\mathbf{k}},\nonumber\\
\longrightarrow&& \frac{\lambda}{N_c}   \sum_{\mathbf{k}'} \sum_{\alpha} m^\ast_{\mathbf{k'},\alpha}m_{\mathbf{k'},\alpha}
m^\ast_{\mathbf{k},\alpha}m_{\mathbf{k},\alpha} A_{\mathbf{k}'} =E_\mathbf{q=0} A_{\mathbf{k}},\nonumber\\
\longrightarrow&& \frac{\lambda}{N_c}   \sum_{\mathbf{k}'} \sum_{\mathbf{k}}\sum_{\alpha} m^\ast_{\mathbf{k'},\alpha}m_{\mathbf{k'},\alpha}
m^\ast_{\mathbf{k},\alpha}m_{\mathbf{k},\alpha} A_{\mathbf{k}'} =E_\mathbf{q=0} \sum_{\mathbf{k}}A_{\mathbf{k}},\nonumber\\
\longrightarrow&& \frac{\lambda}{N_{orb}}   \sum_{\mathbf{k}'}\sum_{\alpha} m^\ast_{\mathbf{k'},\alpha}m_{\mathbf{k'},\alpha}
 A_{\mathbf{k}'} =E_\mathbf{q=0} \sum_{\mathbf{k}}A_{\mathbf{k}}
 ,\nonumber\\
 \longrightarrow&& \frac{\lambda}{N_{orb}}   \sum_{\mathbf{k}'}
  A_{\mathbf{k}'} =E_\mathbf{q=0} \sum_{\mathbf{k}}A_{\mathbf{k}}\longrightarrow E_\mathbf{q=0}=\frac{\lambda}{N_{orb}}. 
\end{eqnarray}
This means Eq.~\eqref{Eq:eigenenergy_separable_potential} is exact for $\mathbf{q}=0$ provided the uniform pairing condition is satisfied. This further motivates the approximation done in deriving equation (9) of the main text.

\subsection{The relation to the quantum metric}

We continue within the uniform pairing condition defined above.
We consider small momenta and expand Eq.~(\ref{Eq:eigenenergy_separable_potential}) to the second order in $\mathbf{q}$ (sum over repeated indices is assumed, and $\partial_i\equiv\partial/\partial k_i$), $|\langle m_{\mathbf{k-q}/2}|m_{\mathbf{k+q}/2}\rangle|^2= |\langle  m_{\mathbf{k}} - \frac{q_i}{2} \partial_i m_{\mathbf{k}}  + \frac{1}{8} q_i q_j \partial_i \partial_j m_{\mathbf{k}}  + \mathcal{O}(q^3)| m_{\mathbf{k}} + \frac{q_i}{2} \partial_i m_{\mathbf{k}}  + \frac{1}{8} q_i q_j \partial_i \partial_j m_{\mathbf{k}}  + \mathcal{O}(q^3)\rangle|^2$, which gives
\begin{eqnarray}
E_\mathbf{q}\approx \frac{\lambda}{N_{orb}}- q_i q_j\frac{\lambda}{2 N_{orb}}\frac{1}{N_c}\sum_{\mathbf{k}}g_{ij}(\mathbf{k}),
\end{eqnarray}
where $g_{ij}(\mathbf{k})$ is the quantum metric that is given in terms of the Bloch function projector $P_{\mathbf{k}}=|m_{\mathbf{k}}\rangle \langle m_{\mathbf{k}} |$ as 
\begin{align}
    g_{ij}(\mathbf{k}) = \mathrm{Tr} \left[ \partial_i P_{\mathbf{k}} \partial_j P_{\mathbf{k}} \right] .
\end{align} 
The effective mass is 
\begin{eqnarray}\label{Eq:Effective_mass_quantum_metric}
\left[\frac{1}{m^\ast}\right]_{ij}=-\frac{\lambda}{N_{orb}}\frac{1}{N_c}\sum_{\mathbf{k}}g_{ij}(\mathbf{k}),
\end{eqnarray}
which is the same as the result from the multiband BCS theory, see Eq.~\eqref{Eq:Effective_mass_BCS} in Sec. \ref{Sec:BCS}, and the same as the result of equation (9) of the main text.

\subsection{The separable potential approximation}

The energy Eq.~\eqref{Eq:eigenenergy_separable_potential} is  the  trace of the approximated potential Eq.~\eqref{Eq:projected_potential_pprox1}, and this inspires us to
approximate the eigenenergy by the trace of the projected potential, i.e.,
\begin{eqnarray}
E_{\mathbf{q}}=
\frac{1}{N_c}\mathrm{Tr}V_{\mathbf{k},\mathbf{k}'}(\mathbf{q})
&=&\frac{\lambda}{N_c} \sum_{\mathbf{k}} \sum_{\alpha} V_\alpha m^\ast_{\mathbf{k+q}/2,\alpha}m_{\mathbf{k-q}/2,\alpha}
m^\ast_{\mathbf{k-q}/2,\alpha}m_{\mathbf{k+q}/2,\alpha},\\
&=&\frac{\lambda}{N_c}\sum_{\mathbf{k}} \sum_{\alpha} V_\alpha \langle \alpha | P_{\mathbf{k+q}/2} |\alpha\rangle \langle \alpha | P_{\mathbf{k-q}/2} |\alpha\rangle. \label{Eq:projected_potential_pprox2}
\end{eqnarray}
Here $P_{\mathbf{k}}=|m_{\mathbf{k}}\rangle \langle m_{\mathbf{k}} |$ is again the projection operator. Taking the trace of the projected potential corresponds to the separable potential approximation done in the derivation presented in the main text and Section I above. 
As a comparison,  Eq.~\eqref{Eq:eigenenergy_separable_potential} can be written in terms of the projection operator as
\begin{eqnarray}
E_{\mathbf{q}}
&=&
\frac{\lambda}{N_c}\frac{1}{N_{orb}}\sum_{\mathbf{k}} \mathrm{Tr} \left[ P_{\mathbf{k+q}/2} P_{\mathbf{k-q}/2} \right] =
\frac{\lambda}{N_c}\frac{1}{N_{orb}}\sum_{\mathbf{k}} \sum_{\alpha,\beta} \langle \beta | P_{\mathbf{k+q}/2} |\alpha\rangle \langle \alpha | P_{\mathbf{k-q}/2} |\beta\rangle.
\end{eqnarray}
\begin{figure}
	 \includegraphics[width=0.8\columnwidth]{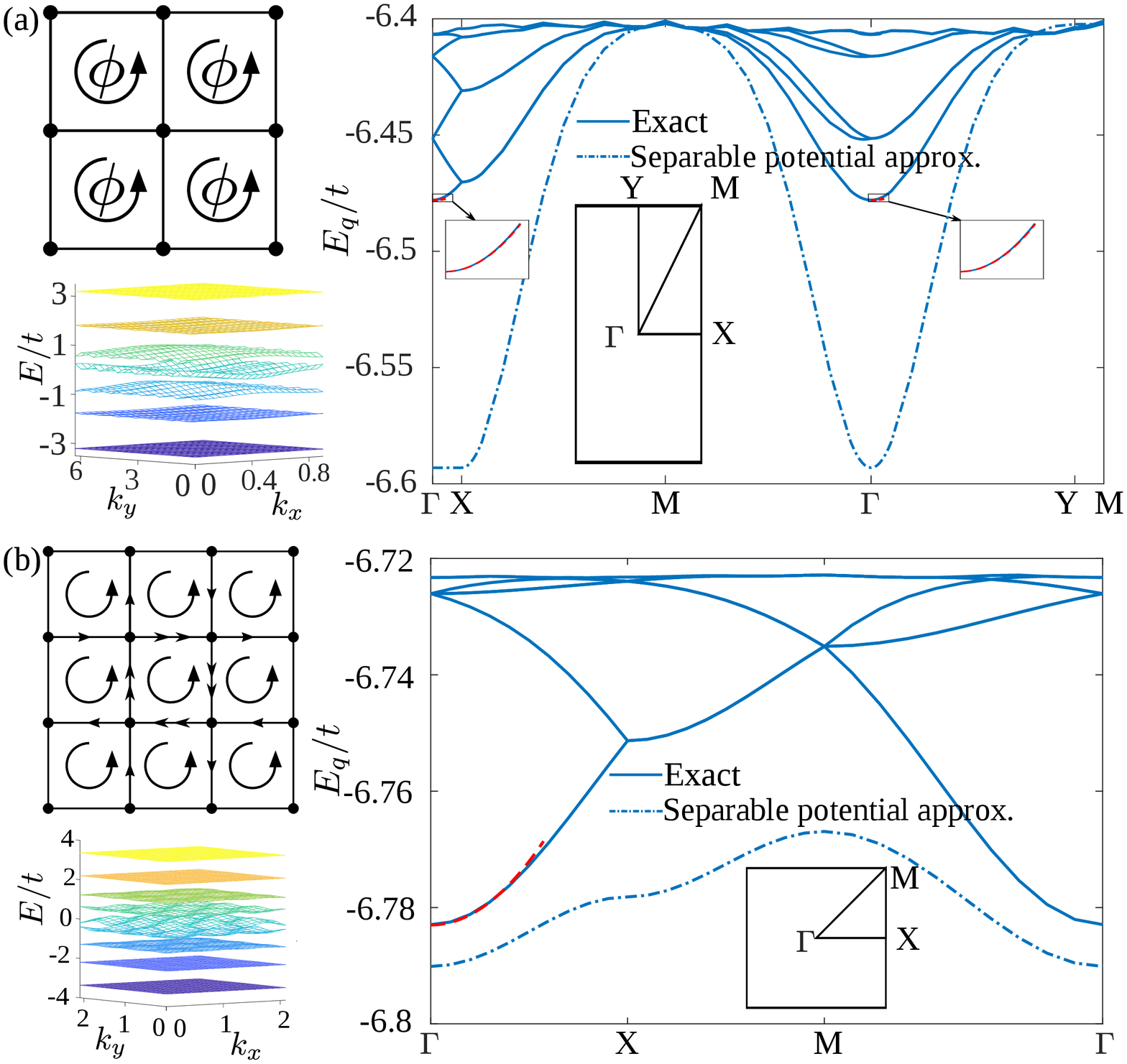}
	\caption{The energy dispersion of  two-body bound states for the time reversal invariant Harper-Hubbard model. 
	The lattice structure and non-interacting energy bands are shown on the left.  The Harper model is defined on a square lattice with nearest-neighbor hopping in the presence of a uniform magnetic field, and has been realized with ultracold atoms \cite{HarperModelBloch,HarperModelKetterle}. The hopping amplitude is $t$ and the magnetic flux through a plaquette is $\phi=2\pi/n_{\phi}$. The magnetic field has opposite signs for opposite spin components, such that the time reversal symmetry is retained. In (a) we take  $n_\phi=7$ and use a gauge where the hopping along the $x$-direction is real,  and therefore the momentum $\mathbf{k}$ is defined in the range $-\pi/(n_\phi a)\le k_x< \pi/(n_\phi a)$ and $-\pi/a \le k_y < \pi/a$ with $a=1$ being the lattice constant. In (b) we take $n_{\phi}=9$ such that we can use a symmetric gauge. A spin up particle acquires  a phase $e^{i\pi/9}$ ($e^{i 4\pi/9}$) when hopping along the links with a single (double) arrow.   The interaction strength is taken to be $\lambda=-0.5 t$. Right: The solid blue curves are the exact solutions of Eq.~(\ref{Eq:boundstates}) and the dashed the separable potential result Eq.~\eqref{Eq:projected_potential_pprox2}. The red curves show the quadratic dispersion with the effective mass given by the integrated quantum metric, Eq.~\eqref{Eq:Effective_mass_quantum_metric}, which agree very well with the exact result for small momentum. }\label{Fig:two_body_dispersion}
\end{figure}
Expanding Eq.~\eqref{Eq:projected_potential_pprox2} for small $\mathbf{q}$, we find
 \begin{eqnarray}
 E_{\mathbf{q}}
 &=&\frac{\lambda}{N_c}\sum_{\mathbf{k}} \sum_{\alpha} V_\alpha \langle \alpha | P_{\mathbf{k+q}/2} |\alpha\rangle \langle \alpha | P_{\mathbf{k-q}/2} |\alpha\rangle,\\
 &=&\frac{\lambda}{N_c}\sum_{\mathbf{k}} \sum_{\alpha} V_\alpha \langle \alpha | P_{\mathbf{k}}+\frac{q_\mu}{2}\partial_\mu P_{\mathbf{k}}+\frac{q_\mu q_\nu}{8} \partial_\mu \partial_\nu P_{\mathbf{k}} |\alpha\rangle \langle \alpha | P_{\mathbf{k}}-\frac{q_\rho}{2}\partial_\rho P_{\mathbf{k}}+\frac{q_\rho q_\lambda}{8} \partial_\rho \partial_\lambda P_{\mathbf{k}} |\alpha\rangle,\nonumber\\
 &=&E_{\mathbf{q=0}}+\frac{q_i q_j}{4}\frac{\lambda}{N_c}\sum_{\mathbf{k}} \sum_{\alpha} V_\alpha \bigg(
 \langle \alpha | P_{\mathbf{k}} |\alpha\rangle \langle \alpha| \partial_i \partial_j P_{\mathbf{k}} |\alpha\rangle - 
 \langle \alpha | \partial_i P_{\mathbf{k}} |\alpha\rangle \langle \alpha|  \partial_j P_{\mathbf{k}} |\alpha\rangle
 \bigg),\\
 &=&E_{\mathbf{q=0}}-\frac{q_i q_j}{2}\frac{\lambda}{N_c}\sum_{\mathbf{k}} \sum_{\alpha} V_\alpha 
 \langle \alpha | \partial_i P_{\mathbf{k}} |\alpha\rangle \langle \alpha|  \partial_j P_{\mathbf{k}} |\alpha\rangle. \label{EnergyInSeparablePotentialAppr}
 \end{eqnarray}
Partial integration as explained in Section I was applied in the last step.
The effective mass is
 \begin{eqnarray}\label{Eq:effective_mass_basis_independent}
 \left[\frac{1}{m} \right]_{ij}=-\frac{\lambda}{ N_c}\sum_{\mathbf{k}} \sum_{\alpha} V_\alpha \langle \alpha | \partial_i P_{\mathbf{k}} |\alpha\rangle \langle \alpha|  \partial_j P_{\mathbf{k}} |\alpha\rangle. 
 \end{eqnarray}
This effective mass is the same as equation (8) of the main text, but just expressed in the tight-binding description and the system volume $N_c$ explicitly written out.

The trace of the potential is the total energy of the bound states and in general, there are more than one bound state for a given $\mathbf{q}$, so Eq.~\eqref{Eq:projected_potential_pprox2} not a very good approximation in general.
However, as we have shown numerically in the main text, for the sawtooth ladder and Lieb lattice, the dispersion of the high energy bound state is weak, so Eq.~\eqref{Eq:effective_mass_basis_independent} is a good approximation to the effective mass of the lowest bound state in those models. 

\subsection{The Harper-Hubbard model}

Here we study another model, the time-reversal invariant Harper-Hubbard model. 
Both the exact result and the result of the separable potential approximation are shown in Fig. \ref{Fig:two_body_dispersion}. 
The exact result is obtained by solving the full two-body problem Eq.~(\ref{Eq:boundstates}) numerically.
We plot the separable potential approximation result, Eq.~\eqref{Eq:projected_potential_pprox2}, with a shift given by twice the mean energy of the lowest noninteracting flat band.
Whether the separable potential approximation is good or not depends on the gauge. As can be seen in Fig. \ref{Fig:two_body_dispersion}(a), 
within the separable potential approximation, the dispersion along the $\Gamma-X$ line is flat, which is qualitatively different from the exact result; while along the $\Gamma-Y$ line, the dispersion is non-flat and the approximation becomes better.  It seems that the separable potential approximation becomes much better in a symmetric gauge, as shown in Fig. \ref{Fig:two_body_dispersion}(b). 
The quantum metric approximation to the the effective mass, Eq.~\eqref{Eq:Effective_mass_quantum_metric}, works very well, as can be seen in Fig. \ref{Fig:two_body_dispersion}. The reason might be that the uniform pairing condition is satisfied in this model. 

\section{Effective mass from the multiband BCS result}\label{Sec:BCS}

According to the previous results on the flat band superfluid weight~\cite{Peotta:2015,Liang:2017a}, the change of the energy density $\delta\mathcal{E}/N_c$ in the presence of supercurrent $\mathbf{q}$ is
\begin{eqnarray}
&& \frac{\delta \mathcal{E}}{N_c}=\frac{D_{ij}q_i q_j}{2},\\
&& D_{ij}=\frac{\Delta^2}{2E}\frac{1}{N_c}\sum_{\mathbf{k}}g_{ij}(\mathbf{k}).\label{Eq:SW}
\end{eqnarray}
Here $E=\sqrt{(\epsilon_0-\mu)^2 + \Delta^2}$ ($\epsilon_0$ is the flat band energy) is the energy of the Bogoliubov quasiparticle.
Note that in Refs.~\cite{Peotta:2015,Liang:2017a}, the total momentum of a Cooper pair is $2\mathbf{q}$, while in the current work, we use a different convention such that the total momentum of a Cooper pair is $\mathbf{q}$, and therefore  the superfluid weight, Eq.~\eqref{Eq:SW}, is $1/4$ of the result in the above mentioned publications. 

We apply the BCS result of the order parameter. Since there are multiple bands, the BCS self-consistent equations are (we have assumed that the energy of the flat band is zero and the that pairing is uniform~\cite{Peotta:2015}, namely $\Delta_\alpha = \Delta$ for all orbitals $\alpha$ that participate in the pairing)
\begin{eqnarray}
&& \Delta= \frac{|\lambda|\Delta }{2N_{orb}\sqrt{\mu^2+\Delta^2}},\label{Eq:BCS_selfconsistentDelta}\\
&& n=\frac{1}{2 N_{orb}}\left(1-\frac{\mu}{\sqrt{\mu^2+\Delta^2}}\right).
 \end{eqnarray}  
Here $N_{orb}$ is the number of orbitals, and the density is defined as
\begin{eqnarray}
n N_{orb}=\frac{N_\uparrow+N_\downarrow}{2N_{c}}-\bar{n},
\end{eqnarray}
where $\bar{n}$ is the number of fully occupied non-flat bands. With is convention, the particle number in the flat band is $2N_c N_{orb} n$, and therefore $n_c=n N_{orb}$ can be identified as the density of Cooper pairs in the flat band. Solving the self-consistent equations we find 
\begin{eqnarray}
\frac{\Delta^2}{E}=\frac{2|\lambda| (n_c-n^2_c)}{N_{orb}}.
\end{eqnarray}
So the energy change in the presence of supercurrent is
\begin{eqnarray}
\delta \mathcal{E}=\frac{|\lambda|}{2} \frac{N_c (n_c -n^2_c)}{N_{orb}} \frac{1}{N_c}\sum_{\mathbf{k}}g_{ij}(\mathbf{k}) q_i q_j.
\end{eqnarray}
If there is only one Cooper pair, $N_c (n _c-n^2_c)\approx N_c n_c =1$, then,
\begin{eqnarray}
\delta \mathcal{E}=\frac{|\lambda|}{2 N_{orb}} \frac{1}{N_c}\sum_{\mathbf{k}}g_{ij}(\mathbf{k}) q_i q_j.
\end{eqnarray}
Therefore the effective mass of the Cooper pair is
\begin{eqnarray}\label{Eq:Effective_mass_BCS}
 \left[\frac{1}{m^*}\right]_{ij}=\frac{|\lambda|}{N_{orb}} \frac{1}{N_c}\sum_{\mathbf{k}}g_{ij}(\mathbf{k}).
\end{eqnarray}
Using the effetive mass, the superfluid density can be written as
\begin{eqnarray}
 D_{ij}=n_c(1-n_c)\left[\frac{1}{m^*}\right]_{ij}.  \label{SFWandEffMass}
\end{eqnarray}
The above expression has a clear physical meaning. 
The Cooper pairs are like bosons, and in the low density limit, the superfluid weight should be the density of Cooper pairs divided by the mass of the pair, i.e., $n_c\left[\frac{1}{m^*}\right]_{ij}$, while in the high density limit, it is convenient to use the hole picture so the superfluid weight should be $(1-n_c)\left[\frac{1}{m^*}\right]_{ij}$. The formula~(\ref{SFWandEffMass}) is just an interpolation between these two limits.

\bibliography{ReferencesS}